\documentclass[aps,10pt,pra,twocolumn, superscriptaddress, showpacs, floatfix
]{revtex4-2}

\usepackage{graphicx}
\usepackage{dcolumn}
\usepackage{amsmath, amsfonts}
\usepackage{dsfont}
\usepackage[normalem]{ulem}
\usepackage{amssymb}
\usepackage{hyperref}
\usepackage{times}
\usepackage{braket, physics}
\hypersetup{colorlinks=true,linkcolor=blue,citecolor=blue,urlcolor=blue}
\usepackage{tcolorbox}
\usepackage{booktabs}

\DeclareMathOperator{\arcosh}{arcosh}

\begin{document}

\title{CV-QKD with noisy coherent states and realistic displacement receivers}

\author{L. F. Melo}\email{lucas.felipe@fieb.org.br}
\affiliation{QuIIN - Quantum Industrial Innovation, Centro de Competência EMBRAPII CIMATEC, SENAI CIMATEC, Av. Orlando Gomes, 1845, Salvador, Bahia 41850-010, Brazil}
\author{J. J. Lacerda}
\affiliation{QuIIN - Quantum Industrial Innovation, Centro de Competência EMBRAPII CIMATEC, SENAI CIMATEC, Av. Orlando Gomes, 1845, Salvador, Bahia 41850-010, Brazil}
\author{John A. Mora Rodríguez}
\affiliation{QuIIN - Quantum Industrial Innovation, Centro de Competência EMBRAPII CIMATEC, SENAI CIMATEC, Av. Orlando Gomes, 1845, Salvador, Bahia 41850-010, Brazil}

\author{Alexandre B. Tacla}
\affiliation{QuIIN - Quantum Industrial Innovation, Centro de Competência EMBRAPII CIMATEC, SENAI CIMATEC, Av. Orlando Gomes, 1845, Salvador, Bahia 41850-010, Brazil}

\begin{abstract}
Near-optimal binary coherent state discrimination can be achieved by performing a displacement operation followed by photon detection. This displacement serves as a control parameter that can be tuned to optimize a chosen figure of merit, which makes such receivers attractive beyond state discrimination. Here, we turn this discrimination strategy into a complete continuous-variable quantum key distribution (CV-QKD) protocol based on a binary alphabet of noisy coherent states (displaced thermal states) and displacement receivers. We develop a strategy for estimating the channel parameters with the same displacement receiver, since standard quadrature-based methods do not apply. While a key map can be obtained with on/off detection, we demonstrate that reliable parameter estimation requires photon-number-resolving detection with sufficient resolution. Finite detector resolution underestimates the observed noise and constitutes a potential security loophole, which we quantify and show to be mitigable with modest resolution. We analyze the protocol's security against entangling-cloner collective attacks in the asymptotic regime, for both direct and reverse reconciliation, and show that it can generate secret keys even under noisy state preparation and imperfect detection. Interestingly, we find that the displacement that minimizes the discrimination error probability also performs well for key generation, thus providing a useful and analytically tractable tuning strategy for the CV-QKD receiver. We evaluate the protocol's performance over a range of transmittances and noise parameters, showing that the displacement receiver can outperform conventional homodyne detection in experimentally relevant regimes. Together, our results point to displacement receivers as a promising alternative for CV-QKD implementations.
\end{abstract}

\maketitle

\section{Introduction}

Discrete-modulated continuous-variable quantum key distribution (DM-CV-QKD) encodes key information in a discrete alphabet of Gaussian states, typically coherent states, overcoming the challenges of implementing near-continuous constellations and costly post-processing required by Gaussian-modulated CV-QKD~\cite{Pirandola20, Zhang24, Usenko25, Anka26}. Thanks to its simplicity, DM-CV-QKD has been demonstrated in numerous experimental setups spanning various constellation shapes and sizes~\cite{Roumestan21, Roumestan21b, Pereira22, Pan22, Wang2022, Tian23, Roumestan24, Hajomer24, Hajomer2025, Pan2026}, and even with passive sources that further simplify state preparation by eliminating the need for a modulator and a quantum random number generator~\cite{Zhang26passive}. This ongoing progress is showing that high-performance, long-distance operation is already within reach of current telecommunication technology, even for small alphabets~\cite{Pan2026, Zhang26passive}.

Although coherent states are natural carriers in those settings, thermal states are also a viable source for CV-QKD and have been studied across a range of wavelengths and physical scenarios~\cite{Usenko25, Usenko10, Papanastasiou2018pra, Papanastasiou2021, Filip2008, Weedbrook12}. Their viability has been reinforced by an increasing number of experimental demonstrations exploiting thermal sources for passive state preparation, first in fiber-based demonstrations at metropolitan distances~\cite{Huang2021, Ji2024}, and more recently in free-space field trials~\cite{Yin2025}. Discrete thermal state protocols, however, remain far less explored. Moreover, all of these works rely on conventional Gaussian coherent detection at the receiver, for which security proofs are better established and which integrates naturally with existing telecom infrastructure~\cite{Zhang24}. Here we take a different route, moving beyond the homodyne/heterodyne paradigm. We propose and analyze a two-state CV-QKD protocol in which key information is encoded on displaced thermal states (noisy coherent states) and decoded with a displacement receiver -- a displacement operation followed by on/off detection~\cite{Kennedy73,Takeoka08,Wittmann10}.

Displacement receivers are well-known strategies in the field of quantum state discrimination, being considered for binary modulations of coherent~\cite{Takeoka08, Wittmann10, izumi12, DiMario18, DiMario22} and displaced thermal states~\cite{Yuan20, Zhao21, Kunz22}. These measurement schemes achieve better discrimination than coherent detection (standard quantum limit) for a wide range of parameters, such as amplitude and thermal noise, a feature that is only possible for non-Gaussian measurements~\cite{Roberson21, Moran26}. The displacement is a free control parameter that can be tuned to optimize any chosen figure of merit, which makes these receivers attractive beyond state discrimination.

Despite these advantages, displacement receivers have not yet been established as the basis of a complete CV-QKD architecture. Their use for CV-QKD was first investigated in simplified scenarios, restricted to direct reconciliation~\cite{Wittmann10, Zhao21-QKD, Zhao24, Miao26}, which limits the achievable distance, and to attenuation-only channels with ideal photon detection~\cite{Zhao21-QKD, Zhao24}. None of them, however, provided a strategy to estimate the channel parameters with these receivers, even though it is essential for proving security~\cite{Anka26}. Moreover, reaching long distances over optical fiber requires reverse reconciliation, and realistic implementations are unavoidably affected by detector imperfections and excess noise that can arise from any component of the physical setup. Here we address all these points, establishing a complete and noise-tolerant CV-QKD protocol based on displacement receivers. Since displacement receivers with on/off and photon-number-resolving detection have already been demonstrated for binary coherent-state discrimination~\cite{DiMario18, DiMario22}, both ends of our protocol rely only on already-demonstrated components.

Our main results concern the security of this protocol, its performance, and the estimation of the channel parameters. Specifically, we develop a security analysis of CV-QKD with displacement receivers under entangling-cloner collective attacks in the asymptotic regime, for both direct and reverse reconciliation. Assuming that Alice's and Bob's devices are trusted, we show that secret keys can be generated even under noisy preparation and imperfect detection. We also show that the receiver optimized for quantum state discrimination performs well for key generation, providing a useful and analytically tractable tuning strategy for the CV-QKD receiver. Although noisy preparation limits the secure distance depending on the level of channel noise, the protocol can outperform conventional homodyne detection in experimentally relevant regimes. In addition, we propose a strategy for estimating the quantum channel parameters with the displacement receiver and show that, while a key map can be obtained with on/off detection, reliable parameter estimation with this strategy requires photon-number-resolving detection. Finite detector resolution underestimates the observed noise, a potential security loophole that we quantify and show to be mitigable with modest resolution. 

This paper is organized as follows. In Sec.~\ref{sec:qkd}, we present the physical steps of our protocol, Sec.~\ref{sec:security} establishes the assumptions underlying our security analysis, and in Sec.~\ref{sec:displacement} we discuss the displacement receivers that will be considered in our analysis. In Sec.~\ref{sec:parameter}, we propose a strategy using displacement receivers to estimate the relevant quantum channel parameters and analyze the effects of finite detector resolution. Sec.~\ref{sec:results} presents a performance analysis of the protocol for different regimes of noise. Sec.~\ref{sec:conclusion} presents our concluding remarks and future perspectives.

\section{Description of the protocol}\label{sec:qkd}

In this section, we present a CV-QKD protocol using a binary alphabet of single-mode displaced thermal states and displacement receivers in direct or reverse reconciliation. For completeness, we also briefly introduce displaced thermal states and the positive operator-valued measure (POVM) representation of the displacement receiver with realistic photon detectors. Together, these elements form the building blocks of our protocol.

\textbf{(i) State preparation:} In each round of the protocol, Alice randomly prepares, with equal probabilities, a displaced thermal state in the set $\{\hat{\rho}_\textrm{th}(\alpha_k, \bar{n}_\textrm{A}):= \hat{D}(\alpha_k) \hat{\rho}_\textrm{th}(\bar{n}_\textrm{A}) \hat{D}^{\dagger}(\alpha_k)\}_{k=0}^1$, where $\hat{D}(\alpha_k)=\exp(\alpha_k \hat a^\dag - \alpha_k^* \hat a)$ is the displacement operator in phase space, with $\hat{a} (\hat a^\dag)$ being the annihilation (creation) operator and $\alpha_k=(-1)^k\alpha$ with $\alpha \in\mathbb{R}^+$. Here, $\hat{\rho}_\textrm{th}(\bar{n}_\textrm{A})$ represents a thermal state with mean photon number $\bar{n}_\textrm{A}\in\mathbb{R}^+$, which in the Fock basis $\{\ket{n}\}$ is given by
\begin{equation} \label{eq:thermal}
    \hat\rho_\textrm{th}(\bar{n}_\textrm{A}) = \sum_{n=0}^{\infty} \frac{\bar{n}_\textrm{A}^{n}}{(\bar{n}_\textrm{A}+1)^{n+1}} \, |n\rangle\langle n|.
\end{equation}
In Appendix~\ref{App:thermalstate}, we give a formal derivation of the matrix elements of $\hat{\rho}_\textrm{th}(\alpha_k, \bar{n}_\textrm{A})$ in Fock basis.

In phase space, displaced thermal states can also be conveniently described by their Glauber-Sudarshan $P$-representation~\cite{Glauber63}:
\begin{align} \label{eq:thermalstate}
    \hat{\rho}_\textrm{th}(\alpha_k,\bar{n}_\textrm{A}) = \int_{\mathbb{C}} \textrm{d}^2c ~ \frac{e^{-\frac{|c-\alpha_k|^2}{\bar{n}_\textrm{A}}}}{\pi \bar{n}_\textrm{A}} |c\rangle\langle c|,
\end{align}
where $|c\rangle=\hat D(c)\ket{0}$ is an arbitrary coherent state. Displaced thermal states have mean photon number $\Tr[\hat{n} \,\hat{\rho}_\textrm{th}(\alpha_k,\bar{n}_\textrm{A})]= \alpha^2 + \bar{n}_\textrm{A}$, where $\hat{n}=\hat{a}^\dagger\hat{a}$. To avoid confusion, from here on, we refer to $\alpha^2$ and $\bar{n}_\textrm{A}$ as the mean numbers of signal and thermal photons, respectively. Note that by setting $\bar{n}_\textrm{A}=0$, we recover a binary phase-shift keying coherent state alphabet, since $\lim_{\bar{n}_\textrm{A} \rightarrow 0} \hat{\rho}_\textrm{th}(\alpha_k,\bar{n}_\textrm{A}) = |\alpha_k\rangle\langle\alpha_k|$. For this reason, displaced thermal states are often referred to as noisy coherent states, and may be used as a simple yet insightful model for investigating the effect of noise in state preparation~\cite{Usenko10, Kunz22, Wu24}.

\textbf{(ii) Measurement:} Bob receives the states after transmission through an insecure quantum channel and performs the displacement operation $\hat{D}(\beta)$, where $\beta\in\mathbb{R}^+$, followed by an on/off photon detection. An off-detection event is used to infer that the state prepared by Alice was $\hat{\rho}_\textrm{th}(-\alpha, \bar{n}_\textrm{A})$, whereas on-detection identifies the state $\hat{\rho}_\textrm{th}(\alpha,\bar{n}_\textrm{A})$. The displacement receiver is then characterized by the POVM elements
\begin{subequations}
\begin{align}
    &\hat{\Pi}_{1} = \hat{D^{\dagger}}(\beta) \hat{\Pi}_{\text{off}} \hat{D}(\beta), \label{eq:POVM1}\\
    &\hat{\Pi}_{0} = \mathds{1} - \hat{\Pi}_{1}, \label{eq:POVM2}
\end{align}
\end{subequations}
where $\mathds{1}$ is the identity operator. Here, $\hat{\Pi}_{\text{off}}$ belongs to the POVM $\{\hat{\Pi}_{\text{on}},\hat{\Pi}_{\text{off}}\}$ of realistic on/off detection~\cite{izumi12, DiMario18, DiMario22, notarnicola23}:
\begin{subequations}
\begin{align}
    &\hat{\Pi}_{\text{off}} = e^{-\nu} \sum_{n=0}^\infty(1-\gamma)^n|n\rangle\langle n|, \label{eq:Pioffnonideal}\\
    &\hat{\Pi}_{\text{on}} = \mathds{1} - \hat{\Pi}_{\text{off}}, \label{eq:Pionnonideal}
\end{align}
\end{subequations}
where $\gamma\in[0,1]$ is the quantum efficiency and $\nu\in\mathbb{R}^+$ is the mean number of dark counts. Sub-unit quantum efficiency has the effect of attenuating the signal amplitude, whereas dark counts are spurious detection events, typically described by Poisson statistics~\cite{izumi12}. Ideal detection corresponds to $\nu=0$ and $\gamma=1$, i.e. $\hat{\Pi}_\text{off} =|0\rangle\langle 0|$. Note that the displacement $\beta$ serves as a control parameter over the on/off detection probabilities, which can be tuned, for instance, to maximize the receiver's performance. The optimal choice of the displacement $\beta$ strongly depends on the chosen figure of merit~\cite{Kunz22}. In Sec.~\ref{sec:displacement}, we present different strategies that can be adopted by Bob and show that this choice only impacts the performance of the protocol, rather than its security under the assumptions presented in Sec.~\ref{sec:security}.

\textbf{(iii) Parameter estimation:} After a few rounds of preparation and measurement, Alice and Bob randomly select a subset of data that will be used to estimate the channel parameters. This subset of data must be announced and henceforth discarded. To do this, in each round $j\in\mathbb{N}$, Alice associates the prepared state with an element of the set $\mathcal{A}=\{a_0,a_1\}$ according to the rule:
\begin{align}\label{eq:sincronization}
    \mathcal{A} = 
    \begin{cases}
        a_0~~\text{ if }~\hat{\rho}_j= \hat{\rho}_\textrm{th}(\alpha,\bar{n}_\textrm{A}),\\
        a_1~~\text{ if }~\hat{\rho}_j= \hat{\rho}_\textrm{th}(-\alpha, \bar{n}_\textrm{A}) .
    \end{cases}
\end{align}

As we show in detail in Sec.~\ref{sec:parameter}, our estimation strategy requires photon-number-resolving (PNR) detection. Bob therefore records the number of detected photons in every round, which is coarse-grained into on/off events in the key generation rounds. For the selected rounds, Alice announces her labels and Bob his photon-number outcomes, from which they estimate the channel parameters.

\textbf{(iv) Key map and information reconciliation:} Bob performs a key map that associates his measurement's results with an element of the set $\mathcal{B}=\{b_0,b_1\}$ to obtain the raw key. By performing an on/off detection, Bob obtains a fixed detector response whenever one or more photons are detected (`on' event), and no response in the absence of photons (`off' event). In each round $j\in\mathbb{N}$, the traditional rule for building the raw key is as follows:
\begin{align}\label{eq:decisionrule}
    \mathcal{B} = 
    \begin{cases} 
        b_0~~\text{if }~ \text{`on' } (n_j\geq 1), \\
        b_1~~\text{if }~ \text{`off' } (n_j=0) .
    \end{cases}
\end{align}

In direct reconciliation, Alice sends information to Bob through an authenticated classical channel to correct his string bits. In reverse reconciliation, Bob is the one that sends information to Alice. The details of this step are not within the scope of this paper. For more details, see~\cite{luo2024overview, Usenko25}.

\textbf{(v) Error correction and privacy amplification:} Alice and Bob choose suitable error correction and privacy amplification protocols to generate a secret key.

\section{Security analysis} \label{sec:security}

In this section, we establish the assumptions underlying our security analysis. This includes theoretical modeling of the quantum channel and the trustfulness of Alice and Bob's devices.

\subsection{Asymptotic secret key rate}

QKD security relies on the requirement that the information shared by the trusted parties exceeds the information accessible to the eavesdropper. This requirement is quantified by the secret key rate, which in the asymptotic limit and under collective attacks is determined by the Devetak-Winter formula~\cite{Devetak05}:
\begin{align}  \label{eq:keyrate}
    R^\infty = \mu_\textrm{rec}~I(A;B) - \chi(E;Y),
\end{align}
where $\mu_\textrm{rec}\in[0,1]$ is the reconciliation efficiency~\cite{yang2023information}, $I(A;B)=H(A)-H(A|B)$ is the classical mutual information between Alice and Bob, with $H$ being the Shannon entropy, and $A$ and $B$ are the random variables representing preparation and measurement, respectively~\cite{Cover06}; and $\chi(E;Y)$ is the Holevo bound on the eavesdropper's information on Alice's preparation ($Y=A$, direct reconciliation) or on Bob's measurement outcomes ($Y=B$, reverse reconciliation)~\cite{Holevo73}.

For two equiprobable states, the classical mutual information is given by
\begin{align}  \label{eq:mutualstandard}
    I(A;B) = 1 - \sum_{k=0}^1 p(b_k) H(\lbrace p(a_\ell|b_k) \rbrace),
\end{align}
where $p(b_k) = \frac{1}{2}\sum_{\ell = 0}^{1} p(b_k|a_\ell)$ is the probability of Bob obtaining the result $b_k$, and $p(a_\ell|b_k)$ is the probability that Alice sent $a_\ell$ given that Bob's measurement outcome was $b_k$, which is given by the Bayes rule:
\begin{align} \label{eq:Bayes}
    p(a_\ell|b_k) = \frac{p(b_k|a_\ell)}{2p(b_k)},
\end{align} 
where $p(b_k|a_\ell)=\Tr(\hat{\Pi}_k \hat{\rho}_\textrm{B}^\ell)$ with $\hat{\rho}_\textrm{B}^\ell$ being the state received by Bob after transmission through the quantum channel if Alice prepared the state $\hat{\rho}_\textrm{th}(\alpha_\ell,\bar{n}_\textrm{A})$.

The Holevo bound on the eavesdropper's information on Alice's preparation (direct reconciliation) is given by
\begin{align} \label{eq:HolevoA}
    \chi(E;A) = S(\hat{\rho}_\textrm{Eve}) - \frac{1}{2} \sum_{\ell=0}^1 S(\hat{\rho}_{\textrm{Eve}|a_\ell}),
\end{align}
with states $\hat{\rho}_\textrm{Eve}$ and $\hat{\rho}_{\textrm{Eve}|a_\ell}$ being determined by Eq.~\eqref{eq:EveconditionalDR}. Whereas the Holevo bound on the eavesdropper's information on Bob's measurement outcomes (reverse reconciliation) is given by
\begin{align} \label{eq:HolevoB}
    \chi(E;B) = S(\hat{\rho}_\textrm{Eve}) - \sum_{k=0}^1 p(b_k) S(\hat{\rho}_{\textrm{Eve}|b_k}),
\end{align}
with states $\hat{\rho}_\textrm{Eve}$ and $\hat{\rho}_{\textrm{Eve}|b_k}$ being determined by Eq.~\eqref{eq:Eveconditional}.

The mutual information shared by Alice and Bob with the displacement receiver can be calculated for this scenario using the probabilities provided in Sec.~\ref{sec:displacement}. However, the Holevo bound on Eve's information cannot be further simplified, requiring numerical calculation by truncating the Hilbert space on a suitable number of photons (more details in Sec.~\ref{sec:results}).

Any noise in the protocol reduces the mutual information between Alice and Bob. The same noise affects the Holevo bound on Eve's information about Alice and Bob's raw key in very different ways, thus having a crucial role in the security of CV-QKD. Typically, this bound cannot be determined analytically. To circumvent this problem, one may derive upper bounds through Gaussian extremality properties or numerical optimizations over all channels compatible with the observed statistics~\cite{Ghorai19, Lin19, Denys21}. Closer to our setting, finite constellations of displaced thermal states have been analyzed within composable security proofs against Gaussian collective attacks, but relying on conventional heterodyne detection~\cite{Papanastasiou2021}. However, the non-Gaussian character of the on/off detection considered here makes such a worst-case, channel-model-independent analysis considerably more involved. As a first approach, we assume a realistic physical model of a Gaussian quantum channel. In this way, we can derive Eve's states directly from the channel model and calculate the information leakage without resorting to Gaussian approximations or computationally intensive optimization problems, enabling a preliminary assessment of the protocol's performance  for realistic noise parameters. A more sophisticated security analysis against the worst-case channel is left for future work.

\subsection{Eve's physical attack and channel model}\label{sec:noisy}

\begin{figure}[t]
\centering
\includegraphics[width=\columnwidth]{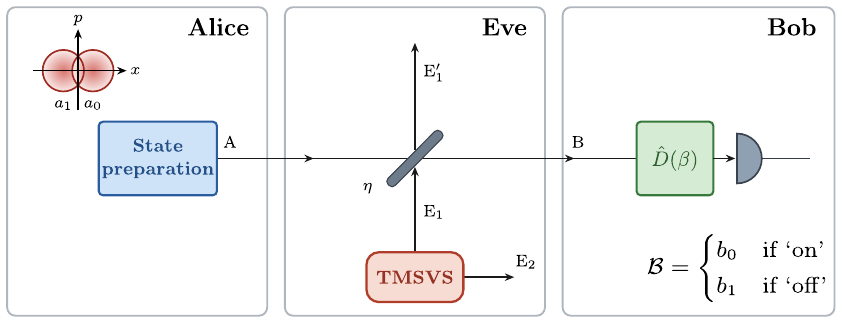}
\caption{Prepare and measure scheme of the protocol under entangling-cloner collective attack. Alice prepares mode $\textrm{A}$ and sends it through the quantum channel. Eve couples Alice's mode to mode $\textrm{E}_1$ of a two-mode squeezed vacuum state (TMSVS) on a beam-splitter with transmissivity $\eta\in[0,1]$, which corresponds to the channel transmissivity. Eve keeps modes $\textrm{E}'_1$ and $\textrm{E}_2$ in a quantum memory for future measurements. The output mode $\textrm{B}$ is sent to Bob, who performs a displacement operation followed by on/off detection.} \label{fig:scheme}
\end{figure}

The Gaussian quantum channel can be modeled as a thermal-loss channel, in which the signal sent by Alice is mixed in a beam splitter with an appropriate thermal state~\cite{grosshans2003virtual, grosshans2007continuous, papanastasiou2018quantum,Usenko25}. However, this model only reproduces the action of the channel and does not provide sufficient conditions to evaluate the security of the protocol. To address this issue, the channel model must include an additional system containing all potentially leaked information from the protocol implementation, which is assumed to be under Eve's control. Under this security assumption, the channel can be equivalently modeled as the result of an entangling-cloner attack performed by Eve (see Fig.~\ref{fig:scheme}). In this attack, Eve mixes one mode of a two-mode squeezed vacuum state (TMSVS) with Alice's mode at a beam splitter with transmissivity that corresponds to the channel attenuation. Eve then forwards one of the beam-splitter outputs to Bob while retaining her two modes. Under collective attacks, she stores these modes in a quantum memory for subsequent measurements.

The TMSVS can be written as
\begin{align}
    |\textrm{TMSV}\rangle_\textrm{E$_1$E$_2$} = (1-\lambda^2) \sum_{k,\ell=0}^\infty \lambda^{k+\ell} |k\rangle_\textrm{E$_1$}\langle\ell|\otimes|k\rangle_\textrm{E$_2$}\langle\ell|,
\end{align}
where $\lambda=\tanh[1/2\arcosh(2\bar{n}_\textrm{E}+1)]$ with $\bar{n}_\textrm{E}$ being the mean number of photons in each of Eve's thermal modes. The global state after the beam-splitter operation in the entanglement-based (EB) scheme~\cite{Usenko25,Anka26} becomes
\begin{small}
\begin{align}
    \hat{\rho}_\textrm{A$_1$BE'$_1$E$_2$} = \hat{B}(\eta) \left( |\Phi\rangle_\textrm{A$_1$A$_2$}\langle\Phi| \otimes |\textrm{TMSV} \rangle_\textrm{E$_1$E$_2$}\langle\textrm{TMSV}| \right)  \hat{B}^\dagger(\eta),
\end{align}
\end{small}

\noindent where $\hat{B}(\eta)= \mathds{1}_\textrm{A$_1$}\otimes \exp[\arccos(\sqrt{\eta})(\hat{a}_\textrm{A$_2$}^\dagger \hat{a}_\textrm{E$_1$}-\hat{a}_\textrm{A$_2$} \hat{a}_\textrm{E$_1$}^\dagger)]\otimes \mathds{1}_\textrm{E$_2$}$ is the beam-splitter operator with transmissivity $\eta$, $\mathds{1}_\textrm{A$_1$}$ and $\mathds{1}_\textrm{E$_2$}$ are the identity operators for modes $\textrm{A$_1$}$ and $\textrm{E$_2$}$, respectively, and Alice holds the purification $|\Phi\rangle_\textrm{A$_1$A$_2$}=(\mathds{1}_\textrm{A$_1$}\otimes\tau^{1/2})\sum_{n=0}^\infty |n\rangle|n\rangle$ with $\tau$ being the average state of the constellation~\cite{Denys21,Mora2026}. In this way, Bob receives the average state $\hat{\rho}_\textrm{B} = \Tr_\textrm{AE'$_1$E$_2$}[\hat{\rho}_\textrm{A$_1$BE'$_1$E$_2$}]$, which results in a mixture of displaced thermal states (see Appendix~\ref{App:channelnoise}): 
\begin{align} \label{eq:Bobchan}
    \hat{\rho}_\textrm{B}= \frac{1}{2} \sum_{\ell=0}^1 \hat{\rho}_\textrm{th} \left(\sqrt{\eta}\alpha_\ell, \eta\bar{n}_\textrm{A} + (1-\eta) \bar{n}_\textrm{E}\right).
\end{align}
Therefore, the mean number of thermal photons of the states entering Bob's receiver is given by:
\begin{align} \label{eq:nB}
    \bar{n}_\textrm{B} =\eta \bar{n}_\textrm{A} + (1-\eta) \bar{n}_\textrm{E}.
\end{align}
Note that if $\bar{n}_\textrm{E}=0$, the channel reduces to a pure-loss channel, where Eve mixes Alice's modes with vacuum on the beam-splitter. Since we assume that Alice's devices are trusted and $\bar{n}_\textrm{A}$ is known by the trusted parties, the excess noise is directly related to the channel-added noise. The relation between these quantities is given by expression~\cite[\S III-B-6]{Usenko25}
\begin{align}  \label{eq:excess}
    \xi = \frac{(1-\eta)}{\eta}2\bar{n}_\textrm{E}.
\end{align}
In Table~\ref{tab}, we show a few reference values for $\bar{n}_\textrm{E}$ corresponding to typical values of excess noise in optical fibers~\cite{Jouguet13,Huang16} with lengths $L=50$, $100$, $150$ and $200$ km.

To compute the Holevo bound in direct reconciliation, we first determine Eve's states conditioned on Alice's preparation, which are given by
\begin{small}
\begin{align} \label{eq:EveconditionalDR}
\hat{\rho}_{\textrm{Eve}|a_\ell}= \frac{\Tr_\textrm{A$_1$B} [(\hat{\textrm{P}}^\ell_\textrm{A$_1$}\otimes \mathds{1}_\textrm{B}\otimes \mathds{1}_\textrm{E}) \hat{\rho}_\textrm{A$_1$BE} (\hat{\textrm{P}}^{\ell,\dagger}_\textrm{A$_1$}\otimes\mathds{1}_\textrm{B} \otimes\mathds{1}_\textrm{E})]}{1/2},          
\end{align}
\end{small}

\begin{table}[t] 
\centering
\begin{tabular}{ccccc} 
    \toprule
    {Excess noise} & \multicolumn{4}{c}{Distance} \\
    \cmidrule(lr){2-5}
Eq.~\eqref{eq:excess} & $50$ km & $100$ km & $150$ km & $200$ km \\ 
\midrule
$0.005$ & $2.8\times10^{-4}$ & $2.5\times10^{-5}$ & $2.5\times10^{-6}$ & $2.5\times10^{-7}$ \\

$0.010$ & $5.5\times10^{-4}$ & $5.0\times10^{-5}$ & $5.0\times10^{-6}$ & $5.0\times10^{-7}$ \\

$0.015$ & $8.3\times10^{-4}$ & $7.6\times10^{-5}$ & $7.5\times10^{-6}$  & $7.5\times10^{-7}$ \\

$0.020$ & $1.1\times10^{-3}$ & $1.0\times10^{-4}$ & $1.0\times10^{-5}$ & $1.0\times10^{-6}$ \\
\bottomrule
\end{tabular}
\caption{Mean number of photons in Eve's thermal mode for typical values of excess noise in optical fibers~\cite{Jouguet13,Huang16}.}\label{tab}
\end{table}

\noindent where $\hat{\rho}_\textrm{A$_1$BE}:= \hat{\rho}_\textrm{A$_1$BE'$_1$E$_2$}$, $\mathds{1}_\textrm{B}$ and $\mathds{1}_\textrm{E}= \mathds{1}_\textrm{E'$_1$} \otimes\mathds{1}_\textrm{E$_2$}$ are the identity operators for Bob and Eve's modes, respectively, and $ \hat{\textrm{P}}^\ell_\textrm{A$_1$}$ is the operator representing the projection that Alice must implement in mode $\textrm{A$_1$}$ to send the state $\hat{\rho}_\textrm{th}(\alpha_\ell,\bar{n}_\textrm{A})$ to Bob. Eve's average state is then given by $\hat{\rho}_\textrm{Eve}= \frac{1}{2} \sum_{\ell=0}^1 \hat{\rho}_{\textrm{Eve}|a_\ell}$.

In reverse reconciliation, Eve's states conditioned on Bob's measurement are determined by
\begin{small}
\begin{align} \label{eq:Eveconditional}
\hat{\rho}_{\textrm{Eve}|b_k}= \frac{\Tr_\textrm{A$_1$B} [(\mathds{1}_\textrm{A$_1$}\otimes \hat{\textrm{M}}_\textrm{B}^k\otimes \mathds{1}_\textrm{E}) \hat{\rho}_\textrm{A$_1$BE} (\mathds{1}_\textrm{A$_1$}\otimes\hat{\textrm{M}}_\textrm{B}^{k,\dagger} \otimes\mathds{1}_\textrm{E})]}{p(b_k)},        
\end{align}
\end{small}

\noindent where $\{ \textrm{M}_\textrm{B}^k \}_{k=0}^1$ are the Kraus operators that represent Bob's measurement, i.e. $\hat{\textrm{M}}_\textrm{B}^{k,\dagger} \hat{\textrm{M}}_\textrm{B}^k= \hat{\Pi}_k$. Eve's average state is then given by $\hat{\rho}_\textrm{Eve}=\sum_{k=0}^1 p(b_k)\hat{\rho}_{\textrm{Eve}|b_k}$.

\section{Displacement receiver strategies} \label{sec:displacement}

The displacement receiver allows the redefinition of the initial phase discrimination problem as a discrimination between states with different amplitudes. As mentioned in Sec.~\ref{sec:qkd}, Bob can adopt different strategies for displacement $\beta$, and the optimal value of this quantity crucially depends on the chosen figure of merit~\cite{Kunz22}. We will analyze three strategies for choosing the displacement $\beta$. These scenarios depend on reference values of the channel parameters that are known or can be estimated with reasonable accuracy. For optimization, we consider that Bob knows the noise added by Alice's device in the preparation of the states $\bar{n}_\textrm{A}$, which we assume is trusted. The noise added by the quantum channel, $\bar{n}_\textrm{E}$, is unknown to the trusted parties. Henceforth, we consider the pessimistic scenario in which Bob assumes that the quantum channel is pure, $\bar{n}_\textrm{E}=0$, so his implementation is necessarily sub-optimal for $\bar{n}_\textrm{E}>0$. With this approach, we can investigate the robustness of each strategy to channel noise.

In practical applications, the parameter $\eta$ is also unknown in principle. However, one can use the expected value of this parameter by knowing the characteristics of the physical channel without the presence of an eavesdropper (for example, by using an \textit{Optical Time-Domain Reflectometer}). A verification step can be added where, during parameter estimation (as discussed in Sec.~\ref{sec:parameter}), we compare this guessed transmittance value with the real value obtained from the classical data statistics to verify the deviation of the displacement parameter and assess whether the protocol can proceed or needs to be aborted. Therefore, assuming an attenuation value would not affect the security of the protocol, but rather its performance. Once the value for the proper displacement is obtained, the protocol is implemented.

We start by calculating the detection probabilities, which are essential for the calculation of both mutual information~\eqref{eq:mutualstandard} and the Holevo bound [see Eqs.~\eqref{eq:HolevoB} and \eqref{eq:Eveconditional}]. Under the Gaussian channel assumption, Bob receives a displaced thermal state in each round~\eqref{eq:Bobchan}, as shown in Sec.~\ref{sec:security}. The probabilities that Bob assigns the bit value $1$ when implementing detection operators described by Eqs.~\eqref{eq:POVM1}, \eqref{eq:POVM2}, \eqref{eq:Pioffnonideal} and \eqref{eq:Pionnonideal} will be given by
\begin{align}  
    p(b_1|a_\ell) &= \Tr \left[\hat{\Pi}_\text{1}\hat{\rho}_\textrm{th}(\sqrt{\eta}\alpha_\ell,\bar{n}_\textrm{B})\right] \nonumber\\
    &=\frac{e^{-\nu}e^{-\gamma\frac{(\beta+\sqrt{\eta}\alpha_\ell)^2}{1+\gamma\bar{n}_\textrm{B}}}}{1+\gamma\bar{n}_\textrm{B}}, \label{eq:disp_probs_b1}
\end{align}
where $\bar{n}_\textrm{B}$ is given by Eq.~\eqref{eq:nB}. To derive this expression, we use Eqs.~\eqref{eq:rhomn} and \eqref{eq:Pioffnonideal}, together with the identity $\sum_{n=0}^\infty \lambda^n L_n^{(0)}(-|x|^2) = e^{\lambda|x|^2 /(1-\lambda)}/(1-\lambda)$. The probabilities of Bob assigning the bit value $0$ can be determined by $p(b_0|a_\ell) = 1 - p(b_1|a_\ell)$. 

\textbf{(i) Kennedy receiver:} Bob chooses $\beta= \sqrt{\eta}\alpha$. For this receiver strategy, Bob minimizes the individual probability of error $p(b_0|a_1)$ by displacing the state $\hat{\rho}_\textrm{th}(-\sqrt{\eta}\alpha,\bar{n}_\textrm{B})$ back to the origin, yielding the thermal state $\hat{\rho}_\textrm{th}(0,\bar{n}_\textrm{B})$. In this case, $p(b_0|a_1)=1-e^{-\nu}/(1 + \gamma \bar{n}_\textrm{B})$, which vanishes in the simplest scenario of modulation of coherent states ($\bar{n}_\textrm{A}=0$), a pure-loss channel ($\bar{n}_\textrm{E}=0$) and zero dark counts ($\nu =0$), yielding a deterministic result when Alice sends the state $|-\alpha\rangle$.  

\textbf{(ii) Optimized displacement receiver for secret key rate:} Bob chooses $\beta$ defined by $\partial R^\infty/ \partial\beta|_{\bar{n}_\textrm{E}=0}=0$, which is the most natural choice of optimization strategy for a CV-QKD receiver. Due to the nonlinear nature of the secret key rate, this optimization must be carried out numerically, making it difficult to interpret its physical meaning.

\textbf{(iii) Optimized displacement receiver for error probability:} Bob chooses $\beta$ defined by $\partial P_e/\partial\beta|_{\bar{n}_\textrm{E}=0}=0$, where $P_e$ is the average error probability:
\begin{align}
    P_e &= \frac{1}{2}[p(b_0|a_1)+p(b_1|a_0)] \label{eq:Pe} \\
    &=\frac{1}{2}\left[1-\frac{e^{-\nu- \gamma\frac{(\beta-\sqrt{\eta}\alpha)^2}{1 + \gamma\bar{n}_\textrm{B}}}}{1 + \gamma\bar{n}_\textrm{B}} +  \frac{e^{-\nu-\gamma\frac{(\beta+\sqrt{\eta}\alpha)^2}{1+\gamma\bar{n}_\textrm{B}}}}{1 + \gamma\bar{n}_\textrm{B}} \right]\nonumber\\
    &= \frac{1}{2}-\frac{e^{-\nu- \gamma\frac{(\beta^2+\eta\alpha^2)}{1 + \gamma\bar{n}_\textrm{B}}}}{1+\gamma\bar{n}_\textrm{B}} \sinh \left(\frac{2\gamma \beta\sqrt{\eta}\alpha}{1+\gamma\bar{n}_\textrm{B}} \right). \label{eq:averageerror}
\end{align}
The solution to $\partial P_e/\partial\beta|_{\bar{n}_\textrm{E}=0}=0$ is the transcendental equation:
\begin{align}  \label{eq:betaopt}
    \beta_\textrm{opt}= \sqrt{\eta}\alpha \coth \left(\frac{2\gamma\beta_\textrm{opt}\sqrt{\eta}\alpha}{1+\gamma\eta\bar{n}_\textrm{A}}\right),
\end{align}
which can be solved numerically for a given set of parameters $\{\alpha,\bar{n}_\textrm{A},\eta,\gamma\}$. This is a relevant result for the field of quantum state discrimination, as it generalizes the results found in~\cite{Takeoka08}, which considered coherent states and imperfect detectors, and in~\cite{Yuan20}, which considered displaced thermal states and ideal detectors. It can be immediately seen in Eq.~\eqref{eq:betaopt} that $\beta_\textrm{opt}$ increases with amplitude $\alpha$ and that $\beta_\textrm{opt} \geq \sqrt{\eta}\alpha$; for fixed transmissivity, the equality is obtained asymptotically for high amplitudes ($\alpha \rightarrow \infty$)~\cite{DiMario18, DiMario22}. Note that $\beta_\textrm{opt}$ does not depend on $\nu$, since dark counts are random noise and cannot be prevented by manipulation of the states.

Before comparing the performance of the foregoing receivers, we give an interpretation of the average error probability that can be very insightful. This reasoning, together with the results shown next, provides a connection between CV-QKD and quantum state discrimination. It can be easily shown that the \textit{total variation distance}~\cite{Patrick95} between the probability distributions of obtaining `on' and `off' detections for each state is
\begin{align}  
    d_\textrm{TV} &:= \frac{1}{2}\sum_{k=0}^1 \left|p(b_k|a_0) - p(b_k|a_1)\right| \label{eq:total_variation}\\
    &= 1 - 2P_e, \nonumber
\end{align}
where we used the fact that $p(b_1|a_1)\geq p(b_1|a_0)$ for $\beta\geq0$. Therefore, the displacement that minimizes the error probability is the one that maximizes the average distance between the on/off probability distributions of each state in the signal. In this case, Bob has more confidence on his results reflecting Alice's bits on average. As we show next, this is not the ideal setting for CV-QKD, where more ambiguity in the results is required to reduce Eve's knowledge of the raw key. However, the performance of this strategy is shown to approximate the performance of receiver (ii) depending on the level of channel noise.

\begin{figure}[t]
\centering
    \includegraphics[width=\columnwidth]{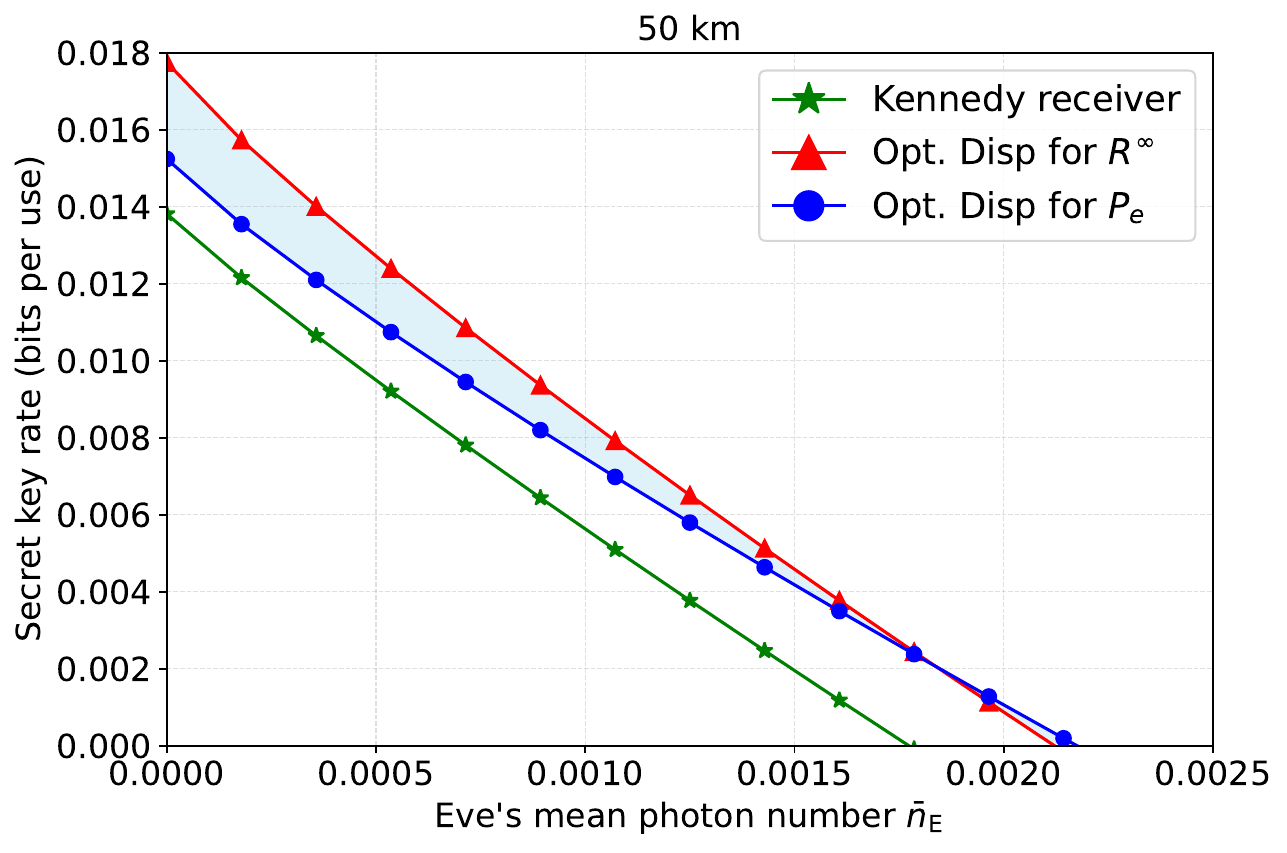}
    \caption{Secret key rate in reverse reconciliation at 50~km as a function of the quantum channel noise for displacement receivers with ideal detectors, assuming that Alice prepares states with $\bar{n}_\textrm{A}=0.05$ and the optimal amplitude for each receiver. The shaded region highlights the difference between the secret key rates of receivers (ii) and (iii), which vanishes at the crossover point $\bar{n}_\textrm{E}\approx1.8\times10^{-3}$.
    } \label{fig:channelnoise}
\end{figure}

Now, we compare the performance of receivers (i), (ii) and (iii). To focus our attention on their tolerance to channel noise, we assume that the detectors are ideal. For all plots, we consider transmission over a commercial optical fiber, i.e. $\eta=10^{-0.02L}$ with $L$ being the distance in kilometers, and assume ideal reconciliation ($\mu_\textrm{rec} = 1$). In Fig.~\ref{fig:channelnoise}, we plot the secret key rate for the three above-mentioned optimized displacement receivers in reverse reconciliation at 50~km as a function of the quantum channel noise parameter $\bar{n}_\textrm{E}$. For this plot, we consider that Alice prepares displaced thermal states with optimized signal amplitude $\alpha$ and $\bar{n}_\textrm{A}=0.05$, which is experimentally feasible~\cite{Zhang26passive}. The optimal amplitude was obtained by numerical grid search over $20$ equally spaced values of $\alpha$ in the interval $[0.4,0.7]$. The Kennedy receiver is outperformed by the other strategies over the entire range of channel noise considered. Although receiver (iii) performs slightly worse than receiver (ii) at low channel noise, it can surpass it at higher noise levels. The crossover point depends on the transmission distance and occurs around $\bar n_\textrm{E}\approx1.8\times10^{-3}$ for $50$ km.

\begin{figure}[t]
\centering
    \includegraphics[width=\columnwidth]{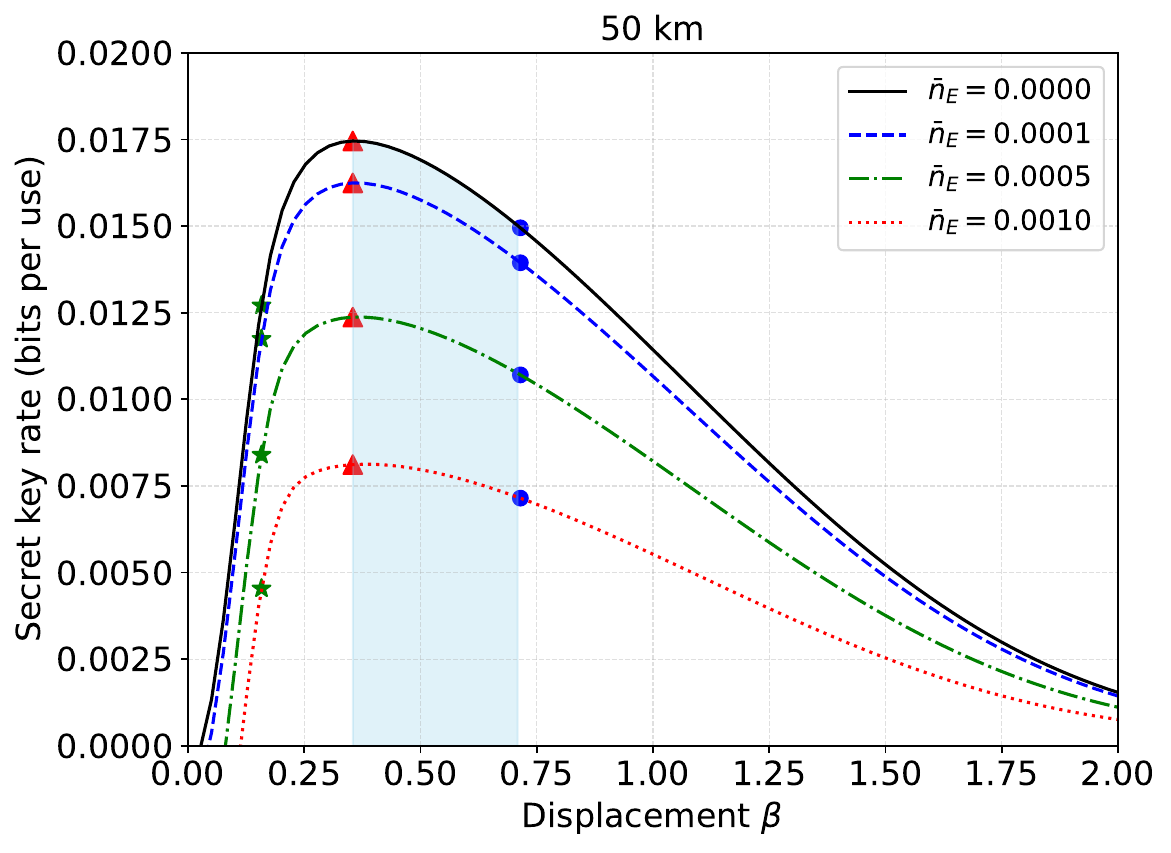}
    \caption{Secret key rate in reverse reconciliation at 50~km as a function of the displacement $\beta$ for different levels of channel noise. The stars, triangles, and circles correspond to the optimization of $\beta$ for receivers (i), (ii) and (iii), respectively. Here, we consider $\alpha = 0.5$, $\bar{n}_\textrm{A}=0.05$ and the values of $\bar{n}_\textrm{E}$ correspond to excess noise $\xi=0, 1.8\times 10^{-3}, 9.0\times 10^{-3}, 1.8\times 10^{-2}$, from top to bottom. The optimal displacement of receiver (ii) ($\beta^{(ii)}_{\textrm{opt}} \approx0.35$) is about half that of receiver (iii) ($\beta^{(iii)}_{\textrm{opt}} \approx0.71$). The interval $\beta\in(\beta^{(ii)}_{\textrm{opt}},\beta^{(iii)}_{\textrm{opt}}]$ (shaded region) corresponds to displacement values that achieve close-to-optimal secret key rates.} \label{fig:KeyvsBeta}
\end{figure}

\begin{figure}[t]
\centering
    \includegraphics[width=\columnwidth]{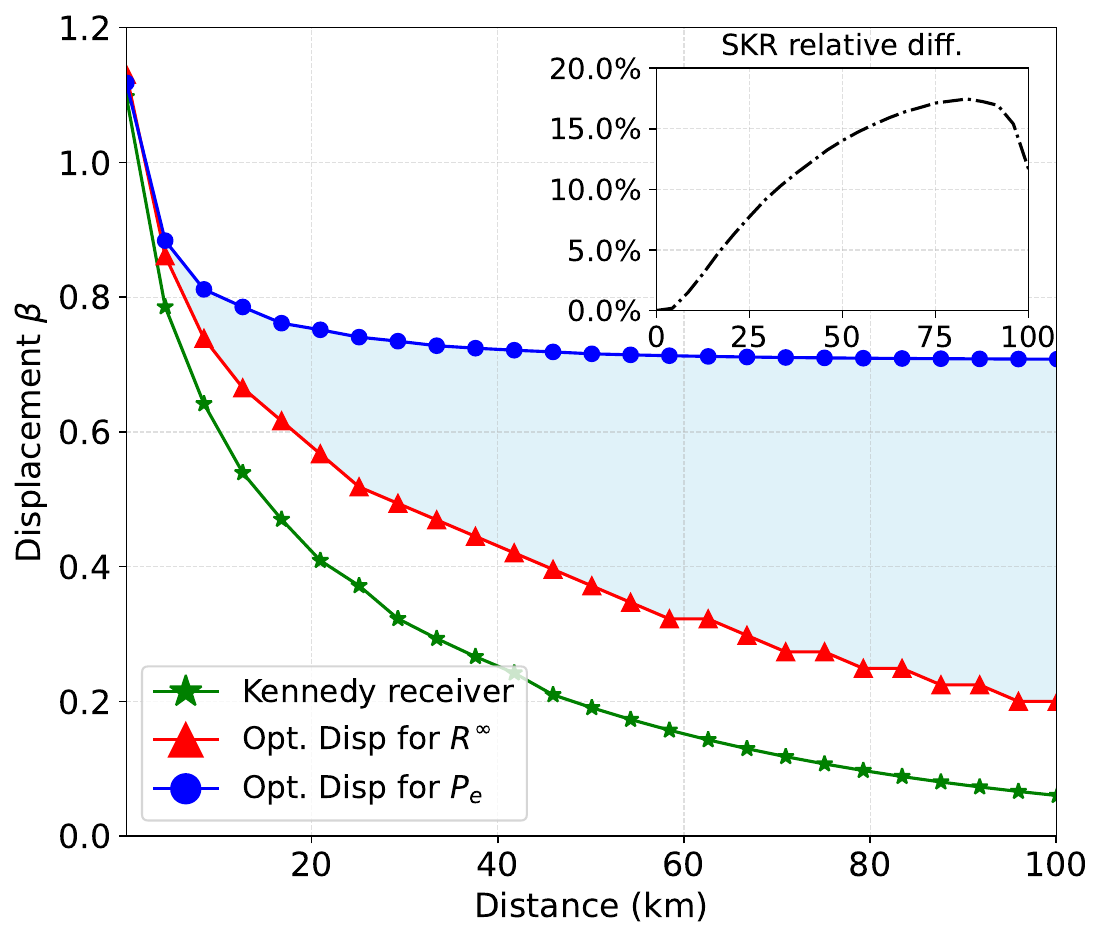}
    \caption{Displacements for receivers (i), (ii) and (iii) as functions of distance for $\bar{n}_\textrm{A}=0.05$ and $\bar{n}_\textrm{E}=10^{-4}$. For each receiver, and at each distance, we use the amplitude that gives maximal secret key rate in reverse reconciliation. Inset shows the relative difference between the secret key rates obtained with the strategies bounding the shaded region. See text for details.} \label{fig:betas}
\end{figure}

Despite the dependence of the protocol's performance on $\beta$, the secret key rate is robust to a wide range of sub-optimal displacements, as shown in Fig.~\ref{fig:KeyvsBeta}, where we show the secret key rate as a function of the displacement $\beta$ for different levels of channel noise. Here, we choose $\alpha=0.5$, $\bar{n}_\textrm{A}=0.05$, and $L=50$ km; stars, triangles, and circles correspond to receivers (i), (ii) and (iii), respectively. Secret keys can be generated only for displacements above a threshold value that increases with the channel noise. The optimal displacement of receiver (ii) ($\beta^{(ii)}_{\textrm{opt}} \approx0.35$) is about half that of receiver (iii) ($\beta^{(iii)}_{\textrm{opt}} \approx0.71$). The shaded region in Fig.~\ref{fig:KeyvsBeta}, $\beta\in[\beta^{(ii)}_{\textrm{opt}},\beta^{(iii)}_{\textrm{opt}}]$, marks the range of displacements over which the secret key rate remains close to its maximum.

\begin{figure*}
\includegraphics[width=\textwidth]{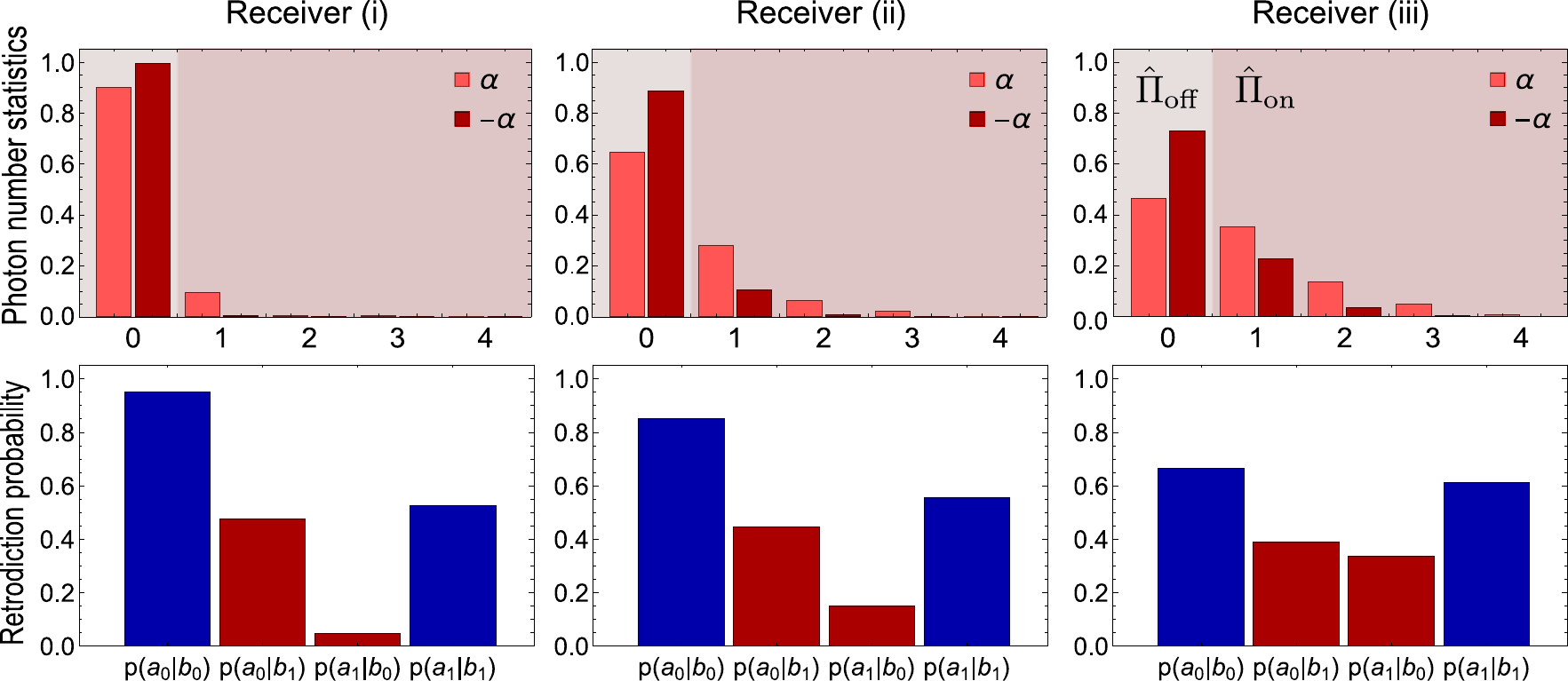}
\caption{Photon number statistics $\langle n|\hat{D}(\beta)\hat{\rho}_\textrm{B}^\ell\hat{D}^\dagger(\beta)|n\rangle$ (top row) and retrodiction probabilities $p(a_\ell|b_k)$ [Eq.~\eqref{eq:Bayes}] (bottom row) for receivers (i) ($\beta^{(i)}_{\textrm{opt}} \approx0.16$), (ii) ($\beta^{(ii)}_{\textrm{opt}} \approx0.35$), and (iii) ($\beta^{(iii)}_{\textrm{opt}} \approx0.71$). We consider the same parameters as Fig.~\ref{fig:KeyvsBeta}, i.e. $\alpha=0.5$, $\bar{n}_\textrm{A}=0.05$, $L=50$ km, with $\bar{n}_\textrm{E}=10^{-3}$. The colored regions in the top row labeled by $\hat{\Pi}_{\textrm{off}}$ and $\hat{\Pi}_{\textrm{on}}$ represent the photon number values corresponding to Bob's decision rule~\eqref{eq:decisionrule}. The optimal scenario for CV-QKD [receiver (ii)] lies in an intermediate scenario between having a high retrodiction $p(a_0|b_0)$ [receiver (i)] and having a minimal average error probability [receiver (iii)]. See text for details.
} \label{fig:statistics}
\end{figure*}

Figure~\ref{fig:KeyvsBeta} shows a specific scenario in which we fix the distance and the amplitude of the signal. To extend our analysis, in Fig.~\ref{fig:betas}, we plot the displacements of receivers (i), (ii) and (iii) as functions of distance using the signal amplitudes that maximize the secret key rate (in reverse reconciliation) of each strategy. The optimal amplitude was found through a numerical grid search with $25$ equally spaced values in the interval $\alpha\in[0.4,1.1]$. For this plot, we set $\bar{n}_\textrm{A}=0.05$ and $\bar{n}_\textrm{E}=10^{-4}$, which corresponds to an excess noise of $\xi=0.02$ at $100$ km (see Table~\ref{tab}). The inset shows that the relative difference between the secret key rates obtained at the boundaries of the shaded region $[\beta^{(ii)}_{\textrm{opt}},\beta^{(iii)}_{\textrm{opt}}]$ remains below $17.4\%$ over the entire range of distances considered. While the displacement given by Eq.~\eqref{eq:betaopt} saturates at $\beta\approx0.71$, the optimal displacements of receivers (i) and (ii) are monotonically decreasing. As a consequence, the interval $[\beta^{(ii)}_{\textrm{opt}},\beta^{(iii)}_{\textrm{opt}}]$ increases with distance, meaning that Bob gains more freedom to choose the value of $\beta$ without compromising much of the secret key rate.

Figures~\ref{fig:KeyvsBeta} and \ref{fig:betas} indicate that optimizing Bob's discrimination of the received states is not necessarily the ideal strategy for secret key generation. However, they also show that the difference between these strategies decreases with the mean number of photons in Eve's thermal mode. In fact, for $50$ km, both strategies have equal performance for $\bar{n}_{\textrm{E}} \approx 1.8 \times 10^{-3}$, i.e., $3.2 \%$ of excess channel noise, which is slightly larger than the typical noise values presented in Table~\ref{tab}. Furthermore, the optimal scenario for quantum state discrimination, given by optimization~\eqref{eq:betaopt}, provides a two-fold advantage in our security analysis: (1) It allows analytical derivation of displacement $\beta$, and (2) it provides lower bounds for secret key rates in typical CV-QKD scenarios (excess noise smaller than $0.02$~\cite{Jouguet13, Huang16, Roumestan21}). For this reason, we only consider receiver (iii) in our analysis of the protocol's performance in Sec.~\ref{sec:results}.

The good performance of receiver (iii) for CV-QKD can be understood more intuitively by addressing the probability distributions of each state. In the top row of Fig.~\ref{fig:statistics}, we show the photon number statistics $\{p(n|a_\ell)\}_{\ell=0}^1$ for $n\in\lbrace 0,\ldots,4\rbrace$ for the same parameters as Fig.~\ref{fig:KeyvsBeta} (i.e., $\alpha=0.5$, $\bar{n}_\textrm{A}=0.05$, $L=50$ km, with $\bar{n}_\textrm{E}=10^{-3}$) and consider from left to right the optimal displacements of receivers (i) ($\beta^{(i)}_{\textrm{opt}} \approx0.16$), (ii) ($\beta^{(ii)}_{\textrm{opt}} \approx0.35$), and (iii) ($\beta^{(iii)}_{\textrm{opt}} \approx0.71$). The colored regions labeled by $\hat{\Pi}_{\textrm{off}}$ and $\hat{\Pi}_{\textrm{on}}$ represent the photon number values corresponding to on/off detection, depicting Bob's decision rule~\eqref{eq:decisionrule} for building the raw key. The bottom row of Fig.~\ref{fig:statistics} presents the retrodiction probabilities $p(a_\ell|b_k)$ [Eq.~\eqref{eq:Bayes}], where the blue (red) bars correspond to correct (erroneous) retrodictions. Receiver (i) minimizes $p(a_1|b_0)$ at the expense of concentrating most of the detection results in `off' events. In this way, it provides high confidence in the retrodiction $p(a_0|b_0)$. To minimize the average probability of error, receiver (iii) distributes retrodictions $p(a_0|b_0)$ and $p(a_1|b_1)$ more fairly, spreading the photon number statistics to maximize Eq.~\eqref{eq:total_variation}. Our results show that neither of these situations is ideal for CV-QKD. In fact, receiver (ii) lies in an intermediate scenario between having a high retrodiction $p(a_0|b_0)$ [receiver (i)] and having a minimal average error probability [receiver (iii)]. Nevertheless, the similarity between the statistics obtained with receivers (ii) and (iii) gives an intuitive view of the results of this section.

\section{Parameter estimation}\label{sec:parameter}

Although secret keys can be generated with on/off detection, the parameter estimation strategy that we propose here requires a photon-number-resolving (PNR) detector. Using PNR detection, Bob can reconstruct the photon number statistics, as shown in Fig.~\ref{fig:statistics}, in the asymptotic limit of measurements, with the estimation accuracy improving as photon-number resolution increases. In parameter estimation, he and Alice reveal part of the collected data to obtain probability distributions. This choice does not affect our calculations of the secret key rates, since Bob performs the decision rule~\eqref{eq:decisionrule} (depicted by the colored regions in Fig.~\ref{fig:statistics}) on the rounds that are designed to generate the raw key -- in this case, PNR detection reduces to on/off detection. 

The POVM elements of a realistic PNR measurement are described by~\cite{barnett98, Izumi13}  
\begin{equation}\label{eq:PNR}
    \hat{\Pi}_n^{\textrm{PNR}} = e^{-\nu} \sum_{l=0}^{n} \sum_{k=n-l}^{\infty} \frac{\nu^l}{l!} \binom{k}{n-l} \gamma^{n-l} (1 - \gamma)^{k-(n-l)} |k\rangle \langle k|.
\end{equation}
on/off reception can be seen as a particular case of the above detection scheme, since $\hat{\Pi}_{\text{off}}= \hat{\Pi}_0^{\textrm{PNR}}$ [Eq.~\eqref{eq:Pioffnonideal}] and $\hat{\Pi}_\textrm{on}= \sum_{n=1}^\infty \hat{\Pi}_n^{\textrm{PNR}}$ [Eq.~\eqref{eq:Pionnonideal}]. To compute the secret key rate, we need a strategy to estimate the relevant channel parameters, i.e. $\bar{n}_\textrm{B}$ and $\eta$, using detection~\eqref{eq:PNR}. Let $Y=\{y_{k}\}_{k=1}^m$ be the subset of Bob's PNR measurement results used for the parameter estimation step. We have
\begin{align}
\bar{y} &:= \frac{1}{m}\sum_{k=1}^{m}y_k \nonumber \\
&=\sum_{n=0}^{\infty}\frac{l_n n}{m}\;\xrightarrow[m\to\infty]{}\; \sum_{n=0}^{\infty}n\,\mathrm{Tr}\!\left[\hat{\rho}_\textrm{B}\hat{D}^\dagger(\beta)\hat{\Pi}_n^{\textrm{PNR}}\hat{D}(\beta)\right], \label{eq:Estimation}
\end{align}
where $l_n$ is the number of times the value $n$ appears in the subset $Y$ and $\hat{\rho}_\textrm{B}$ is given by Eq.~\eqref{eq:Bobchan}. From here on, $\bar{y}$ always refers to the asymptotic limit $m\rightarrow\infty$. After some algebra, we find (for a derivation, see Appendix~\ref{app:PE1}):
\begin{equation} \label{eq:n_estimator0}
\bar{y} \approx \gamma\,\langle\hat{n}\rangle_{\hat{D}(\beta)\hat{\rho}_\textrm{B}\hat{D}^\dagger(\beta)}+\nu.
\end{equation}
Hence, $\bar{y}$ serves as an estimator for $\gamma\,\langle\hat{n}\rangle+\nu$ in state ${\hat{D}(\beta)\hat{\rho}_\textrm{B} \hat{D}^\dagger(\beta)}$. Using the displaced state ${\hat{D}(\beta)\hat{\rho}_\textrm{B}\hat{D}^\dagger(\beta)}$, we obtain:
\begin{equation}\label{eq:n_estimator}
    \langle\hat{n}\rangle=\bar{n}_\textrm{B}+\frac{1}{2}\sum_{\ell=0}^1\left(\sqrt{\eta}\alpha_\ell+\beta\right)^2=\bar{n}_\textrm{B}+\eta\alpha^2+\beta^2,
\end{equation}
which yields the following estimator for $\bar{n}_\textrm{B}$:
\begin{equation}\label{eq:nestimator}
    \bar{n}_\textrm{B}\approx\frac{\bar{y}-\nu}{\gamma}-\eta\alpha^2-\beta^2.
\end{equation}

As can be seen in the above equation, the estimated value of $\bar{n}_\textrm{B}$ still depends on the transmittance $\eta$, which therefore must also be estimated. This can be accomplished using Eq.~\eqref{eq:disp_probs_b1}, by estimating the probabilities $p(b_1|a_k)$ from the measurement data. Specifically, Bob extracts from his parameter estimation data the outcomes that are equal to $b_1$, generating the subset $Y_1=\{y_{k,1}\}_{k=1}^{m_1}\subseteq Y$, which corresponds to the measurement outcomes of the displacement receiver described by the POVM element in Eq.~\eqref{eq:POVM1}. The empirical conditional probabilities are then given by:
\begin{equation}
    \bar{p}(b_1|a_0)=\frac{m_1^{(0)}}{m_1} \hspace{0.5cm} \text{and} \hspace{0.5cm} \bar{p}(b_1|a_1)=\frac{m_1^{(1)}}{m_1},
\end{equation}
where $m_1^{(0)}$ and $m_1^{(1)}$ denote the number of $b_1$ outcomes conditioned on Alice having sent the state associated with $a_0$ and $a_1$, respectively. From these empirical probabilities, we obtain (for a derivation, see Appendix~\ref{app:PE2}):
\begin{equation}\label{eq:estimator}
    \sqrt{\eta}=\frac{(1+\gamma\bar{n}_\textrm{B})A}{4\gamma\beta\alpha},
\end{equation}
where $A:=\ln \left[\bar{p}(b_1|a_1)/\bar{p}(b_1|a_0)\right]$. By substituting the estimator for $\bar{n}_\textrm{B}$ from Eq.~\eqref{eq:nestimator} into Eq.~\eqref{eq:estimator} we find a quadratic equation in $\sqrt{\eta}$ with solutions:
\begin{equation}
    \sqrt{\eta}= \frac{-2\beta\pm \sqrt{4\beta^2+A^2B}}{A\alpha},
\end{equation}
where $B=(1+\bar{y}-\nu)/\gamma-\beta^2$. Since $\bar{p}(b_1|a_1)>\bar{p}(b_1|a_0)$ is expected under normal operating conditions, we have $A>0$, so the positive root is selected. If $\bar{p}(b_1|a_1)<\bar{p}(b_1|a_0)$, Bob is making more errors than correct decisions when discriminating the states. In this case, Eq.~\eqref{eq:estimator} gives:
\begin{equation}
    \frac{4\gamma\beta\sqrt{\eta}\alpha}{1+\gamma\bar{n}_\textrm{B}}=A<0,
\end{equation}
which implies $\beta<0$.\footnote{Provided that the synchronization for the states sent by Alice is correct and follows the relation in Eq.~\eqref{eq:sincronization}.} Since $\beta$ is fixed and chosen to be positive at the start of the protocol, this behavior constitutes strong evidence of eavesdropping (e.g. Eve could intercept the signal and apply a negative displacement to the state that is forwarded to Bob, potentially forcing a swap $a_0\leftrightarrow a_1$). In this case, after verifying whether the labels of the states sent by Alice are being correctly interpreted and that the detector is functioning properly alongside the implementation of the chosen displacement, the protocol must be aborted.

A particularly notable case arises when $\bar{p}(b_1|a_1)=\bar{p}(b_1|a_0)$, i.e., $A=0$. In this case, Bob is not able to distinguish between the two incoming states, which can occur in one of three situations: (i) no displacement is applied ($\beta=0$); (ii) the detector efficiency vanishes ($\gamma\to 0$); or (iii) the channel attenuation is complete ($\eta\to 0$). All three are consistent with $4\gamma\beta \sqrt{\eta}\alpha=0$. Since $\beta$ and $\gamma$ are known and calibrated prior to the protocol, the only viable explanation is $\eta\to 0$, in which case the protocol must also be aborted.

\subsection{Finite resolution detection effect}\label{Finite resolution effect}

The POVM elements of Eq.~\eqref{eq:PNR} assume a PNR detection of infinite resolution, i.e. it can resolve an arbitrary number of photons. However, in realistic implementations PNR detectors present finite resolution, i.e., they resolve up to a fixed number of photons. In this section, we quantify the effect of finite detector resolution on the estimator derived above.

A PNR detector with a resolution of $M\in\mathbb{N}$ photons, PNR($M$), identifies a number of $n$ photons up to $M$ and maps all values $n \geq M$ to $M$. PNR($M$) detection is described by the POVM elements $\{\hat{\Pi}_n^{\textrm{PNR($M$)}}\}_{n=0}^M$ where $\hat{\Pi}^{\textrm{PNR($M$)}}_n := \hat{\Pi}^{\textrm{PNR}}_n$ for $n<M$ and $ \hat{\Pi}^{\textrm{PNR($M$)}}_{M} := \mathds{1} - \sum_{n=0}^{M-1} \hat{\Pi}^\textrm{PNR}_n$.

\begin{figure}[t]
\centering
    \includegraphics[width=\columnwidth]{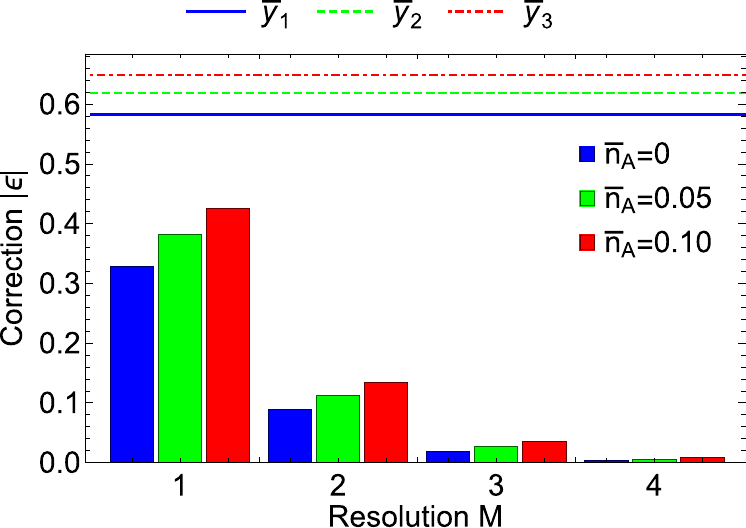}
    \caption{Absolute value of the finite-resolution correction factor [Eq.~\eqref{eq:epsilon}] as a function of the PNR detector resolution, $M$. For the plot, we considered transmission over an optical fiber with length $30$ km and $0.01$ excess noise~[Eq.~\eqref{eq:excess}]. For preparation, we considered the optimal amplitudes of the protocol at the chosen distance for each thermal noise considered: $\alpha_\textrm{opt}\approx 0.53$, $0.59$, $0.63$ for $\bar{n}_\textrm{A}=0$, $0.05$, $0.1$, respectively. We considered detectors with $\gamma=0.85$ and $\nu=10^{-3}$, and displacement according to error probability~\eqref{eq:betaopt}.} \label{fig:epsilon}
\end{figure}

\begin{table}[t] 
\centering
\begin{tabular}{ccccccc} 
    \toprule
    {Distance} &  Estimator & \multicolumn{4}{c}{PNR resolution} \\
    \cmidrule(lr){3-6}
(km) & Eq.~\eqref{eq:n_estimator0} & 1 & 2 & 3 & 4\\ 
\midrule
$50$ & $0.543$ & $0.238$ & $5.5\times10^{-2}$ & $1.0\times10^{-2}$ & $1.5\times10^{-3}$\\

$100$ & $0.504$ & $0.137$ & $2.4\times10^{-2}$ & $3.3\times10^{-3}$ & $3.6\times10^{-4}$\\

$150$ & $0.501$ & $0.116$ & $1.8\times10^{-2}$ & $2.3\times10^{-3}$ & $2.3\times10^{-4}$\\

$200$ & $0.501$ & $0.109$ & $1.7\times10^{-2}$ & $2.0\times10^{-3}$ & $2.0\times10^{-4}$\\
\bottomrule
\end{tabular}
\caption{Absolute values of the correction factor~\eqref{eq:epsilon} for different distances (rows) and PNR resolution (3rd to 6th columns) for excess noise $\xi=0.01$ [see Eq.~\eqref{eq:excess}]. In the 2nd column, we show the value of estimator $\bar{y}$ [Eqs.~\eqref{eq:n_estimator0} and \eqref{eq:n_estimator}]. We considered preparation with $\bar{n}_\textrm{A}=0.05$ and the optimal amplitude of the protocol at each distance, $\alpha_\textrm{opt}\approx0.56$ at $50$ km and $\alpha_\textrm{opt}\approx0.54$ for the other distances.}\label{tab2}
\end{table}

For a PNR($M$), the finite resolution limits Bob's sample space to a maximum of $M$ photons. As a result, if $Y = \{y_k\}_{k=1}^{m}$ is the subset of Bob's measurement used to perform parameter estimation, we have
\begin{align}
    \frac{1}{m}\sum_{k = 1}^m y_k ~ \xrightarrow[m\to\infty]{} &~ \sum_{n = 0}^{\infty} n \Tr\left[\hat{\rho}_\textrm{B}\hat{D}^\dagger(\beta)\hat{\Pi}^\textrm{PNR}_n\hat{D}(\beta)\right] + \epsilon
\end{align}
for the asymptotic limit of measurements, where $\epsilon$ is the correction factor:
\begin{align} \label{eq:epsilon}
    \epsilon = \sum_{n=M}^{\infty} (M-n)\Tr \left[\hat{D}(\beta)\hat{\rho}_\textrm{B}\hat{D}^\dagger(\beta)\hat{\Pi}^\textrm{PNR}_n\right].
\end{align}

\begin{figure}[t]
    \centering
    \includegraphics[width=\columnwidth]{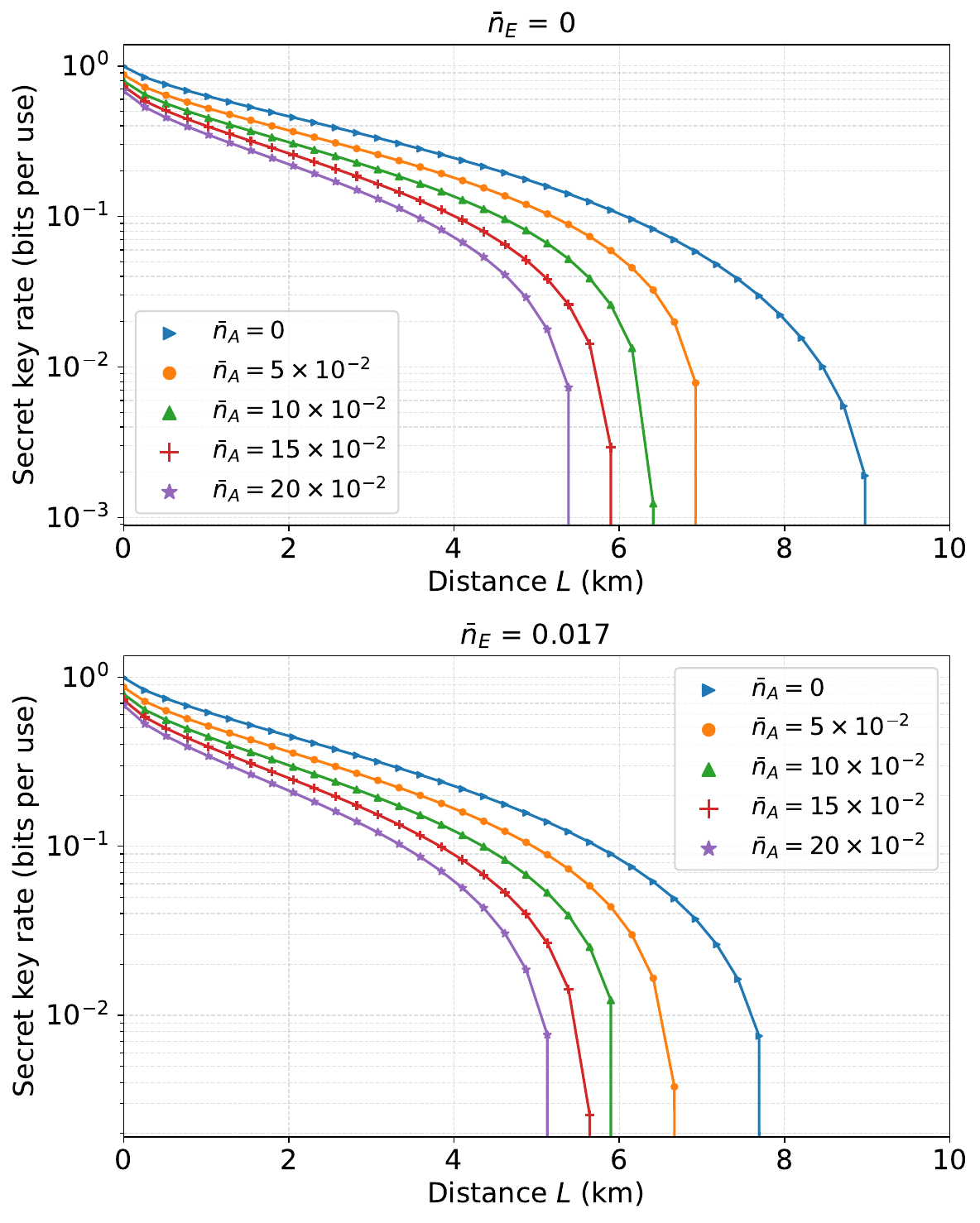}
    \caption{Secret key rates in direct reconciliation with optimized displacement receiver for error probability with realistic detectors ($\gamma=0.85$ and $\nu=10^{-3}$) as functions of distance and preparation noise for the pure-loss channel (top panel) and a thermal-loss channel with $\bar{n}_\textrm{E}=1.7\times10^{-2}$ (bottom panel). This amount of noise corresponds to an excess noise of $0.02$ at distance $10$ km [Eq.~\eqref{eq:excess}].}
    \label{fig:KeyRatesDR}
\end{figure}

\begin{figure*}[t]
\includegraphics[width=\textwidth]{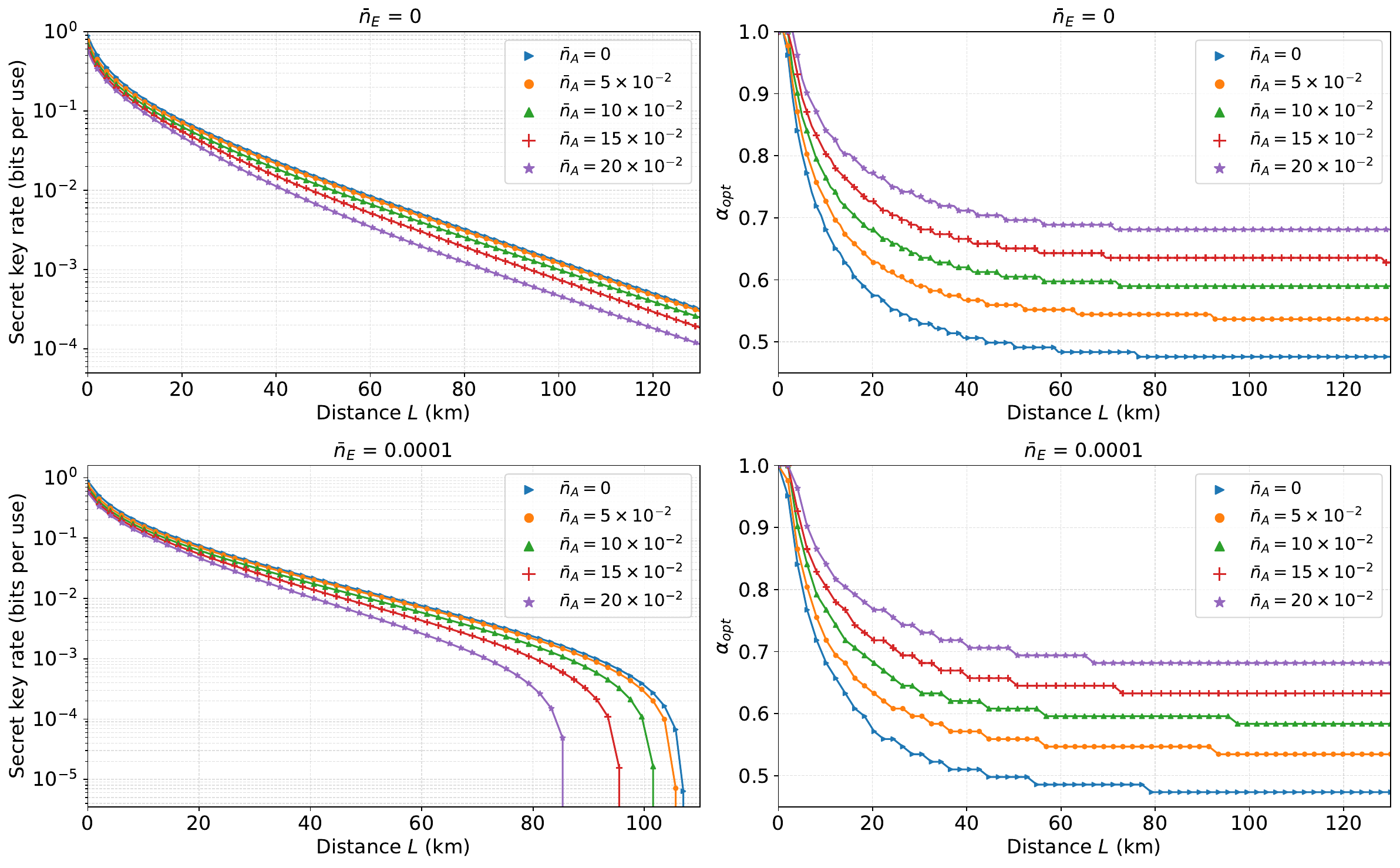}
\caption{CV-QKD with optimized displacement receiver for error probability with realistic detectors ($\gamma=0.85$ and $\nu=10^{-3}$). Secret key rates in reverse reconciliation (left column) and the corresponding optimal signal amplitudes (right column) as functions of distance and preparation noise for the pure-loss channel (top panel) and a thermal-loss channel with $\bar{n}_\textrm{E}=10^{-4}$ (bottom panel). This amount of noise corresponds to an excess noise of $0.02$ at distance $100$ km (Table~\ref{tab}). Both secret key rates and the corresponding optimal amplitudes decrease with attenuation. For the pure-loss channel, preparation noise impacts the protocol's performance while maintaining the viable distance of hundreds of kilometers. For the thermal-loss channel, preparation noise limits the secure distance of the protocol considerably.} \label{fig:KeyRatesRR}
\end{figure*}

This is a negative quantity, which means that the mean number of photons $\bar n_\textrm{B}$ is underestimated for PNR with finite resolution. Naturally, $\epsilon\rightarrow 0$ when $M\rightarrow \infty$, recovering the results of Sec.~\ref{app:PE1}. The underestimation of $\bar n_\textrm{B}$ may constitute a security loophole, however, this loophole can be mitigated with sufficient detector resolution. As shown in Fig.~\ref{fig:epsilon} for 30~km and in Table~\ref{tab2} for longer distances, $|\epsilon|$ decreases rapidly with $M$. Here, we consider implementation with the optimized displacement receiver for error probability [receiver (iii) in Sec.~\ref{sec:displacement}]. For the analysis and throughout this work, we consider realistic detectors implemented at the telecom wavelength ($1550$ nm), the proper choice for transmission over optical fibers. In the literature, quantum efficiencies range from $73\%$ to $98\%$ and dark counts are of the order $\sim10^{-3}$~\cite{Rosenberg05,  Thekkadath21, Stasi23}. To cover this range of parameters, in our simulations we consider detectors with efficiency $\gamma=0.85$ and dark counts $\nu=10^{-3}$. Figure~\ref{fig:epsilon} and Table~\ref{tab2} show the correction factor~\eqref{eq:epsilon} for an excess noise of $\xi=0.01$ [see Eq.~\eqref{eq:excess}] for resolutions $M=1,2,3,4$ (the case $M=1$ corresponds to on/off detection). For the preparation of the states, we considered the optimal amplitudes of the protocol for each thermal noise considered. In Fig.~\ref{fig:epsilon}, $\alpha_\textrm{opt}\approx 0.53$, $0.59$, $0.63$ for $\bar{n}_\textrm{A}=0$, $0.05$, $0.1$, respectively. In Table~\ref{tab2}, where we considered $\bar{n}_\textrm{A}=0.05$, $\alpha_\textrm{opt}\approx0.56$ at $50$ km and $\alpha_\textrm{opt}\approx0.54$ for the other distances.

\section{Performance analysis and discussion}\label{sec:results}

In this section, we present a performance analysis of CV-QKD with the optimized displacement receiver for error probability [receiver (iii), as presented in Sec.~\ref{sec:displacement}] under the security assumptions presented in Sec.~\ref{sec:security}. We consider both direct and reverse reconciliation protocols and compare their performance across different regimes of distance and thermal noise. For all plots, we consider transmission over a commercial optical fiber, i.e. $\eta=10^{-0.02L}$ with $L$ being the distance in kilometers, and assume ideal reconciliation ($\mu_\textrm{rec} = 1$). We calculate the Holevo bound on Eve's information numerically by truncating the Hilbert space at a suitable number of photons. For reverse reconciliation, we employ a truncation of $7$ photons -- for the most extreme values of parameters considered ($\alpha = 1$, $\bar{n}_{\textrm{A}} = 0.2$ and $\bar{n}_{\textrm{E}} = 10^{-4}$), this corresponds to a deviation in the Holevo bound of the order of $10^{-6}$. For the direct reconciliation it is necessary to consider a larger Hilbert space since, in this case, the optimal signal amplitude is usually higher. Thereby, in this scenario, we truncate at $13$ photons which for the values $\alpha = 2$, $\bar{n}_{\textrm{A}} = 0.2$ and $\bar{n}_{\textrm{E}} = 0.017$, provides a deviation in the Holevo bound of the order of $10^{-5}$ compared to using a cut-off of $23$.

In Fig.~\ref{fig:KeyRatesDR}, we plot the secret key rates in direct reconciliation as functions of distance for receiver (iii) with realistic detectors ($\gamma=0.85$ and $\nu=10^{-3}$). For the plots, we choose $\bar{n}_\textrm{A}\in\{0,0.05, 0.10,0.15, 0.20\}$ and transmission through the pure-loss channel (top panel) and a thermal-loss channel (bottom panel) with $\bar{n}_\textrm{E} \approx 1.7\times10^{-2}$, which corresponds to an excess noise of $\xi=0.02$ at distance $10$ km [see Eq.~\eqref{eq:excess}]. To optimize the key rate, we vary the distance in steps of $0.25$ km and perform a numerical search with $80$ values of the amplitude $\alpha$ in the interval $[0.4, 2.0]$. With direct reconciliation, the protocol is limited to short distances, achieving $\sim9$ km for coherent states and pure-loss channel. This limitation is common for direct reconciliation protocols, where the maximum distance (without post-selection) is around $15$ km (3 dB), which can only be achieved with Gaussian modulation of coherent states~\cite{Usenko25, Anka26}. Thermal noise added in preparation reduces the secure distance of the protocol for both the pure-loss and thermal-loss channels.

In Fig.~\ref{fig:KeyRatesRR}, we plot the secret key rates in reverse reconciliation (left column) and the corresponding optimal signal amplitudes (right column) as functions of distance for receiver (iii) with realistic detectors ($\gamma=0.85$ and $\nu=10^{-3}$). We consider different levels of preparation noise and transmission through the pure-loss channel (top panel) and a thermal-loss channel (bottom panel) with $\bar{n}_\textrm{E} =10^{-4}$, which corresponds to an excess noise of $\xi=0.02$ at distance $100$ km (see Table~\ref{tab}). To optimize the key rate, we vary the distance in steps of $2$ km and perform a numerical search with $80$ values of the amplitude $\alpha$ in the interval $[0.4, 1.0]$. Both the secret key rates and the corresponding optimal amplitudes decrease with attenuation. In the upper left panel of the figure, it can be seen that for the pure-loss channel, the protocol can achieve hundreds of kilometers (numerical tests show that one can generate secret keys with rate $\sim10^{-5}$ around $210$ km for $\bar{n}_\textrm{A}$ within the addressed range). In this case, a noisy preparation of the states only degrades the protocol's performance. In the lower left panel, we see that the presence of thermal noise in the quantum channel limits the viable distance of the protocol. The protocol is relatively resilient to the addition of thermal noise in the preparation of the states as long as $\bar{n}_\textrm{A}\sim10^{-2}$. In the case depicted in the bottom row of Fig.~\ref{fig:KeyRatesRR}, this represents a thermal noise that is $100\times$ greater than that of the quantum channel, $\bar{n}_\textrm{E}$. As can be seen in the right column of the figure, to compensate for the presence of thermal noise in the preparation, the signal amplitudes must be increased to achieve optimal protocol performance.

\begin{figure}[t]
\centering
    \includegraphics[width=\columnwidth]{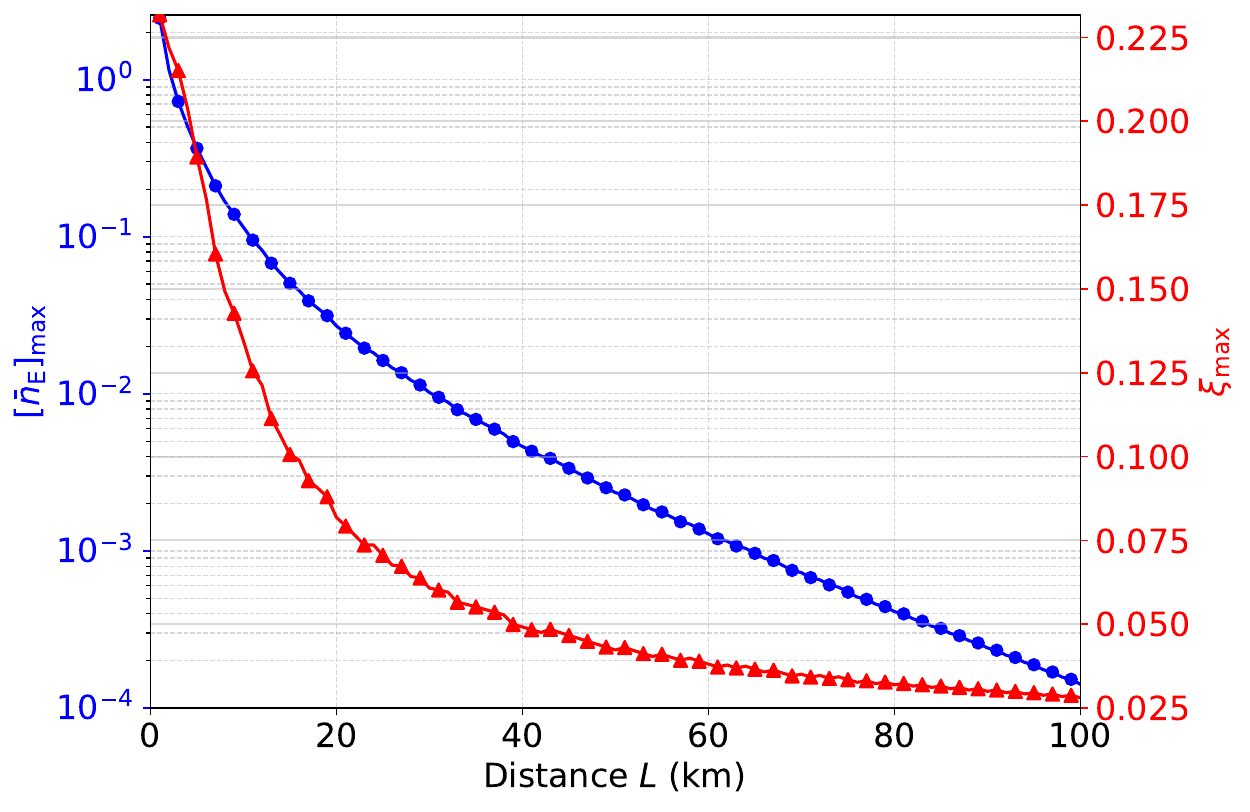}
    \caption{Maximum tolerable mean number of photons in Eve's thermal mode, $[\bar{n}_\textrm{E}]_\text{max}$, (blue circles, left axis) with respect to distance for the protocol in reverse reconciliation. For the plot we consider coherent-state preparation ($\bar{n}_{\textrm{A}} = 0$) since it provides an upper bound for the channel noise. We also show the corresponding excess noise $\xi_\text{max}$ (red triangles, right axis) according to conversion in Eq.~\eqref{eq:excess}.} \label{fig:excess}
\end{figure}

Another important figure of merit of the protocol is the maximum channel noise it tolerates for a given distance. In Fig.~\ref{fig:excess}, we plot the maximum tolerable mean number of photons in Eve's thermal mode, $[\bar{n}_\textrm{E}]_\text{max}$, and the corresponding channel excess noise, $\xi_\text{max}$ [Eq.~\eqref{eq:excess}] as functions of distance, for the protocol in reverse reconciliation. All values were computed for coherent-state preparation ($\bar{n}_{\textrm{A}} = 0$) as it provides an upper bound on the tolerable channel noise. The figure shows that the protocol tolerates a smaller level of noise as distance increases, which is a typical behavior in CV-QKD~\cite{Usenko16, Denys21}.

\begin{figure}[t]
    \centering
    \includegraphics[width=\columnwidth]{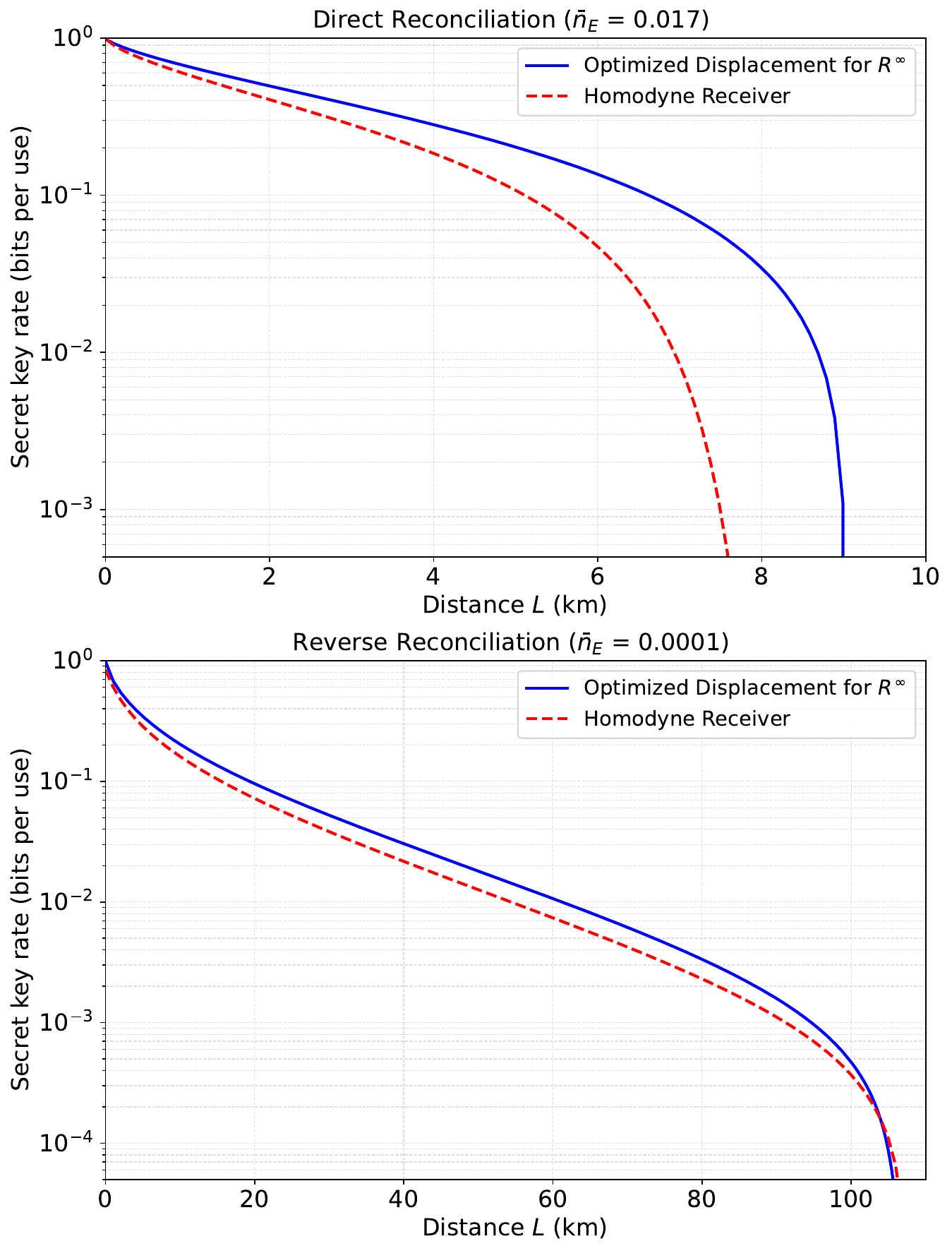}
    \caption{Secret key rate as a function of distance for direct reconciliation (top panel) and reverse reconciliation (bottom panel) for a coherent state alphabet (i.e., $\bar{n}_{\textrm{A}} = 0$) and optimized displacement receiver in comparison to conventional homodyne detection. Both receivers were considered to have ideal detectors. The amplitude of the displacement is chosen to maximize the secret key rate [receiver (ii)]. Here, we considered $\bar{n}_\textrm{E}=1.7\times10^{-2}$ (top) and $\bar{n}_\textrm{E}=10^{-4}$ (bottom), which correspond to an excess noise of $0.02$ at distances $10$ km and $100$ km, respectively [Eq.~\eqref{eq:excess}]. The optimized displacement receiver outperforms conventional homodyne detection for distances $\lesssim 100$ km.}
    \label{fig:KeyRatesHomoVsDisp}
\end{figure}

To explore the full potential of the displacement receiver for CV-QKD, we now consider optimization of the displacement to maximize the secret key rate [receiver (ii), as presented in Sec.~\ref{sec:displacement}]. In Fig.~\ref{fig:KeyRatesHomoVsDisp}, we plot the secret key rates obtained with receiver (ii) and with the homodyne receiver -- both in the cases of ideal detectors. We considered coherent-state preparation ($\bar{n}_{\textrm{A}} = 0$) for both direct reconciliation with channel noise $\bar{n}_{\textrm{E}} = 1.7 \times 10^{-2}$ (top panel) and reverse reconciliation with channel noise of $\bar{n}_{\textrm{E}} = 10^{-4}$ (bottom panel). This amount of noise corresponds to an excess noise of $0.02$ at distances $10$ km and $100$ km, respectively. For the homodyne receiver, the Holevo bound was computed using a truncation of $7$ photons in the reverse reconciliation case. In direct reconciliation, we considered a photon-number cut-off of $13$ for secret key rate below $R^{\infty} = 10^{-3}$, and increased it to $20$ photons for the smaller rates. This adjustment is required because the optimal signal amplitude rapidly increases as the secret key rate falls below $R^{\infty} \approx 10^{-3}$, meaning that the Holevo bound numerical evaluation must account for a larger portion of the Hilbert space. As we can notice, receiver (ii) performs better for both reconciliation scenarios. In particular, for direct reconciliation, this superior performance results in both a higher secret key generation rate and a longer achievable secure distance (a gain of $\sim1.5$ km). For reverse reconciliation, the optimized displacement receiver surpasses the homodyne receiver in secret key rate. In this scenario, the viable secure distance is virtually the same for both receivers, although the homodyne extends marginally farther ($\sim 0.7$ km).

\section{Conclusion and perspectives}\label{sec:conclusion}

In this paper, we present a security analysis of CV-QKD with displacement receivers under entangling-cloner collective attacks in the asymptotic regime. We consider both direct and reverse reconciliation and show that secret keys can be generated even under noisy preparation and imperfect detection. In particular, we show that while a noisy preparation only degrades the performance of the protocol for the pure-loss channel, it limits the viable distance for the thermal-loss channel. Nevertheless, the optimized displacement receiver achieves higher secret key rates than conventional homodyne detection and, in direct reconciliation, a longer secure distance. We further find that the receiver that minimizes the discrimination error probability performs well for key generation, providing a useful and analytically tractable way to choose the displacement receiver for CV-QKD. Moreover, we showed that similar performances can be obtained within a range of displacements, giving Bob freedom to chose his tuning strategy without compromising security. We also develop a strategy for estimating the channel parameters and show that while a key map can be obtained with on/off detection, reliable parameter estimation with this strategy requires photon-number-resolving detection. Finite detector resolution underestimates the observed noise and constitutes a potential security loophole, which we quantify and show to be mitigable with modest resolution. 

We acknowledge that imposing a specific channel model is a restrictive assumption; however, it serves two essential purposes within the protocol. First, it provides physical insight into the sub-optimality of the displacement parameter used in the optimized displacement receiver. Second, and more important, it enables a parameter estimation procedure, which is an indispensable step for any CV-QKD security proof. The study of quantum channel models is widely adopted in the literature as a natural starting point for the development of new protocols and security-proof techniques~\cite{Leverrier11, Zhang24, Usenko25}. We follow the same approach in the initial assessment of CV-QKD protocols based on unconventional detection strategies such as displacement receivers, with the aim of achieving improved performance in terms of secret key rate.

An important research direction is the development of security proofs for arbitrary quantum channels, which remains an open problem for the displacement receiver. In this more general setting, the arbitrariness of the channel introduces significant challenges. In particular, the parameter estimation procedure must be reconsidered, since the current approach strongly relies on the Gaussian channel model. We are currently investigating these directions, which we will present elsewhere. It is also worth investigating whether the discrete outcomes of the displacement receiver can simplify the classical post-processing relative to coherent detection. Taken together, our results position displacement receivers as a promising and experimentally accessible alternative for CV-QKD, while opening the way to further research.

\section*{Acknowledgments}

The authors thank Vladyslav Usenko for insightful discussions and valuable suggestions. This work has been fully funded by the projects ``Receptores não-convencionais em CV-QKD'', ``Análise de segurança de protocolos P$\&$M CV-QKD'' and ``Design e arquiteturas de protocolos CV-QKD'' supported by the EMBRAPII CIMATEC Competence Center in Quantum Technologies-Quantum Industrial Innovation (QuIIN), with financial resources from the PPI IoT/Manufatura 4.0 of the MCTI Grant No. 053/2023, signed with EMBRAPII.

\appendix

\section{Displaced thermal states' density matrix elements}\label{App:thermalstate}

In this appendix, we derive the explicit expression for the matrix element $[\hat{\rho}_\text{th}(\alpha_k,\bar{n})]_{mn}:= [\hat{\rho}_\text{th}^{(k)}]_{mn}$ for arbitrary $\bar{n}$. In Fock space, these states are represented by the expansion:
\begin{align} \label{eq:rho_mnintegral}
    \hat{\rho}_\text{th}(\alpha_k,\bar{n}) = \sum_{m,n=0}^\infty \int \textrm{d}^2c~\frac{e^{-\frac{|c -\alpha_k|^2}{\bar{n}}}}{\pi \bar{n}} e^{-|c|^2} \frac{(c^*)^n c^m}{\sqrt{m!n!}} |m\rangle\langle n|.
\end{align}
By writing the integral of Eq.~\eqref{eq:rho_mnintegral} in polar coordinates, we find
\begin{align}
[\hat{\rho}_\text{th}^{(k)}]_{mn} = \int_0^\infty \textrm{d}r~\frac{e^{-r^2}}{\pi \bar{n}}\frac{r^{n+m+1}}{{\sqrt{m!n!}}} \int_0^{2\pi} \textrm{d}\theta e^{-\frac{|re^{i\theta} -\alpha_k|^2}{\bar{n}}+i(m-n)\theta}. \nonumber
\end{align}
Note that $|re^{i\theta}-\alpha_k|^2=(r\cos\theta-\alpha_k)^2 + r^2\sin^2\theta = r^2+\alpha^2-2\alpha_k r\cos\theta$. We can thus solve the integration on $\theta$ by acknowledging that it will become:
\begin{small}
\begin{align}
    \int_0^{2\pi} \textrm{d}\theta e^{\frac{2\alpha_k r}{\bar{n}}\cos\theta}  \cos[(m-n)\theta] = \left\{ \begin{array}{c} 2\pi I_{(m-n)}\left(\frac{2\alpha_k r}{\bar{n}} \right),\text{ if } m\geq n \displaystyle\\[2mm]
    2\pi I_{(n-m)}\left(\frac{2\alpha_k r}{\bar{n}} \right), \text{ if } m\leq n  \end{array} \right.,\nonumber
\end{align}
\end{small}

\noindent where $I_{k}(x)$ is the Bessel function of the first kind of $x$ and of order $k$:
\begin{align}  \label{eq:Bessel}
    I_{k}(x) = \sum_{\ell=0}^\infty \frac{1}{\ell!(\ell+k)!} \left(\frac{x}{2}\right)^{2\ell+k}.
\end{align}
The coefficient $[\hat{\rho}_\text{th}^{(k)}]_{mn}$ with $m\geq n$ is then given by:
\begin{align}
    \frac{2e^{-\frac{\alpha^2}{\bar{n}}}}{\bar{n}\sqrt{m!n!}} \int_0^\infty \textrm{d}r~e^{-\frac{(1+\bar{n})}{\bar{n}}r^2}r^{n+m+1} I_{(m-n)}\left(\frac{2\alpha_k r}{\bar{n}}\right). \nonumber
\end{align}
Using the definition of the Bessel function~\eqref{eq:Bessel}, and the integral:
\begin{align}
    \int_0^\infty \textrm{d}r~e^{-\frac{(1+\bar{n})}{\bar{n}}r^2}r^{2\ell+2m+1} = \frac{(\ell+m)!}{2} \left(\frac{\bar{n}}{1+\bar{n}}\right)^{\ell+m+1}, \nonumber
\end{align}
we find after some algebra:
\begin{widetext}
\begin{align}
    [\hat{\rho}_\text{th}^{(k)}]_{mn}
    &= \sqrt{\frac{n!}{m!}} \left(\frac{\alpha_k}{\bar{n}}\right)^{m-n} \frac{\bar{n}^{m} e^{-\frac{\alpha^2}{\bar{n}}}}{(1+\bar{n})^{m+1}}
    \sum_{\ell=0}^\infty \frac{1}{\ell!} \left(\begin{array}{c} \ell+m \\ n\end{array} \right) \left[\frac{\alpha^2}{\bar{n}(1+\bar{n})}\right]^{\ell}\nonumber \\
    &= \sqrt{\frac{n!}{m!}} \left(\frac{\alpha_k}{\bar{n}}\right)^{m-n} \frac{\bar{n}^{m} e^{-\frac{\alpha^2}{(1+\bar{n})}}}{(1+\bar{n})^{m+1}}
     \left(\begin{array}{c} m \\ n\end{array} \right) {}_1F_1\left(-n,1+m-n,\frac{-\alpha^2}{\bar{n}(1+\bar{n})} \right).
\end{align}
\end{widetext}
Here, we used the definition of the binomial coefficient $\left(\begin{array}{c} p \\ q \end{array}\right)=\frac{p!}{q!(p-q)!}$, and the Kummer confluent hypergeometric function ${}_1F_1$. Then the identity
\begin{align}
    \left(\begin{array}{c} n+k \\ n\end{array} \right) {}_1F_1\left(-n,1+k,x \right) = L^{(k)}_n (x),
\end{align}
where $L_k^{(l)}(x)$ is the associated Laguerre polynomial of $x$ of degree $k$~\cite{Cariolaro10, Yuan18}, leads to (with $k=m-n$):
\begin{align}  \label{eq:rhomn}
[\hat{\rho}_\text{th}^{(k)}]_{mn} = 
\frac{\bar{n}^m e^{-\frac{\alpha^2}{1+\bar{n}}}}{(1+\bar{n})^{m+1}}\sqrt{\frac{n!}{m!}} \left(\frac{\alpha_k}{\bar{n}}\right)^{m-n}  L_n^{(m-n)}\left[\frac{-\alpha^2}{\bar{n}(1+\bar{n})} \right]
\end{align}

\noindent for $m\geq n$. Matrix elements with $n<m$ can be found using the Hermitian symmetry $[\hat{\rho}_\textrm{th}^{(k)}]_{mn}=[\hat{\rho}_\textrm{th}^{(k)}]_{nm}^*$.

\section{Quantum channel with added thermal noise}\label{App:channelnoise}

In Sec.~\ref{sec:noisy}, we considered a quantum channel with added thermal noise that is controlled by Eve. Our description of this channel consists of Eve performing a beam-splitter mixing of the states sent by Alice with the thermal state $\hat{\rho}_\textrm{E} = \hat{\rho}_\textrm{th}(\bar{n}_\textrm{E})$ [Eq.~\eqref{eq:thermal}]. The beam-splitter operation $\hat{U}_{\text{BS}}(\eta)$, with $\eta\in[0,1]$ being the transmissivity, acts by inducing coupling between two incoming modes \cite{barnett02}. 

To obtain the outgoing modes it is convenient to work with the dynamics in phase space, where Gaussian states evolution are described by suitable transformations of its quadratures first and second moments. In this formalism the first and second moments are characterized, respectively, by the first moments vector (FM) and the covariance matrix (CM) of the states, and unitary transformations by corresponding symplectic transformation \cite{Serafini17,Weedbrook12review}. The FM and CM of an arbitrary displaced thermal state $\hat{\rho} = \hat{\rho}_{\textrm{th}}(z,\bar{n})$ with $\bar{n}$ thermal photons and centered in $z \in \mathds{C}$ is given, respectively, by
\begin{align}\label{eq:cm_thermal_state}
\mathbf{R} = \begin{pmatrix}
 \Tr(\hat{\rho}~\hat{x}_1) \\
 \Tr(\hat{\rho}~\hat{x}_2)
\end{pmatrix}, \quad 
\sigma = \begin{pmatrix}
 \frac{1}{4}(1 + 2 \bar{n}) & 0 \\
0 &  \frac{1}{4}(1 + 2 \bar{n})
\end{pmatrix},
\end{align}
where the covariance matrix elements are obtained through $\sigma_{ij} = \frac{1}{2} \Tr[\hat{\rho}\{\hat{x}_i - R_i, \hat{x}_j - R_j \}]$ with $\hat{x}_{1(2)}$ being the quadrature operator $\hat{q}(\hat{p})$ and $\{,\}$ the anticommutator. Here we defined the quadrature operators as $\hat{q} = \frac{1}{2}(\hat{a} + \hat{a}^{\dagger})$ and $\hat{p} = \frac{1}{2}(\hat{a} - \hat{a}^{\dagger})$, which provides $N_0 = \frac{1}{4}$ for the shot-noise variance.

The symplectic transformation corresponding to a beam splitter of transmittance $\eta$ is given by the $4 \times4$ matrix
\begin{align}
S_{\textrm{BS}}(\eta) = 
    \begin{pmatrix}
        \sqrt{\eta} \mathds{1}_{2} & \sqrt{1 - \eta} \mathds{1}_{2} \\
        -\sqrt{\eta} \mathds{1}_{2} & \sqrt{\eta} \mathds{1}_{2}
    \end{pmatrix}.
\end{align}
The output FM and CM is given by the transformation law
\begin{align}
    \mathbf{R}^{\textrm{out}} = S_{\textrm{BS}}(\eta) \mathbf{R}^{\textrm{in}}\quad \textrm{and} \quad\sigma^{\textrm{out}} = S_{\textrm{BS}}(\eta)\sigma^{\textrm{in}}S^{\mathsf{T}}_{\textrm{BS}}(\eta),
\end{align}
where $\mathbf{R}^{\textrm{in}} (\sigma^{\textrm{in}})$ is the two mode $\mathbf{R}_{\textrm{A}}\oplus \mathbf{R}_{\textrm{B}} ~(\sigma_{\textrm{A}}\oplus \sigma_{\textrm{B}})$ first moment vector (covariance matrix). Here we consider one of the modes as Alice's displaced thermal states $\hat{\rho}_\textrm{A}^\ell= \hat{\rho}_{\textrm{th}}(\alpha_\ell, \bar{n}_{\textrm{A}})$, where $\alpha_\ell=(-1)^\ell\alpha$, with $\ell=0,1$ and $\alpha \in \mathds{R}^{+}$, and the other incoming mode as Eve's thermal state $\hat{\rho}_\textrm{E} = \hat{\rho}_\textrm{th}(\bar{n}_\textrm{E})$. 

Performing the beam splitter transformation and tracing out Eve's output mode, we find for Alice's FM and CM
\begin{align}
\mathbf{R}_{\textrm{A}}^{\textrm{out}} = \begin{pmatrix}
 \sqrt\eta \alpha_{\ell}\\
 0
\end{pmatrix}, \quad \sigma^{\textrm{out}}_{\textrm{A}} &= \frac{1}{4} \left(1 + 2\bar{n}_\textrm{B}\right)\mathds{1}_2,
\end{align}
where $\bar{n}_\textrm{B} = \eta \bar{n}_{\textrm{A}} + (1 - \eta)\bar{n}_{\textrm{E}}$. Comparing with Eq.~\eqref{eq:cm_thermal_state}, we see that the output mode corresponds to the displaced thermal state $\hat{\rho}_\textrm{B}^\ell= \hat{\rho}_{\textrm{th}}(\sqrt{\eta}\alpha_\ell, \eta \bar{n}_{\textrm{A}} + (1 - \eta)\bar{n}_{\textrm{E}})$.

\section{Secret key rate of CV-QKD with binary noisy coherent states and ideal homodyne detection}\label{app:homodyne}

The probability of obtaining the result $x\in\mathbb{R}$ of an ideal homodyne detection implemented by Bob in the incoming states, after propagation through a quantum channel with transmittance $\eta$ is given by
\begin{align}
    p(x|\alpha_k) = \sqrt{\frac{2}{\pi}} \frac{e^{-2\frac{(x-\sqrt{\eta}\alpha_k)^2}{1+2\bar{n}_\textrm{B}}}}{\sqrt{1+2  \bar{n}_\textrm{B}}}.
\end{align}
If $x>0$ ($x<0$), Bob assigns that Alice's bit is $a_0$ ($a_1$). Then, the average probability of error yields
\begin{align}
    P^\textrm{hom}_e &=\frac{1}{2}\left[\int_0^\infty \textrm{d}x~ p(x|a_1) + \int_{-\infty}^0 \textrm{d}x~ p(x|a_0) \right] \nonumber \\
    &= \frac{1}{2} \textrm{Erfc}\left(\frac{\sqrt{2\eta}{\alpha}}{\sqrt{1+2\bar{n}_\textrm{B} }}\right), \label{eq:PeHomodyne}
\end{align}
where $\textrm{Erfc}$ is the complementary error function. 

To compute the secret key rates obtained with the homodyne receiver, we need to calculate the mutual information and the Holevo bound on Eve's information. Unlike the displacement receiver, homodyne detection provides equally probable outputs $p(b_k)=1/2$ for $k=0,1$. Therefore, the mutual information of this strategy yields
\begin{align}
    I(A;B) = 1-\mathcal{H}_2(P_e^\textrm{hom}),
\end{align}
where $\mathcal{H}_2(x)=-x\log_2 x-(1-x)\log_2(1-x)$ is the binary entropy. For direct reconciliation, the Holevo bound is determined using Eq.~\eqref{eq:EveconditionalDR}, so mutual information is the only different ingredient when analyzing different receiver strategies. For reverse reconciliation, the Holevo bound for the homodyne receiver is calculated using Eq.~\eqref{eq:Eveconditional} but substituting the POVM elements of Bob's measurement by $\hat{\Pi}^\pm_\textrm{B} = \int_{x\in\mathbb{R}^\pm} \textrm{d}x~|x\rangle\langle x|$. Using the fact that the outcomes of the homodyne receiver are equiprobable, we find for reverse reconciliation:
\begin{align}
    \chi(E;B) = S(\hat{\rho}_\textrm{E}) - \frac{1}{2} \sum_{k=0}^1 S(\hat{\rho}_{\textrm{E}|b_k}).
\end{align}

\section{Parameter estimation}\label{app:PE}

\subsection{Large-sample limit of PNR measurements}\label{app:PE1}
 
For $\nu=0$, the POVM elements in Eq.~\eqref{eq:PNR} are given by
\begin{equation}
    \Pi_n^{\textrm{PNR}}=\sum_{k=n}^\infty \begin{pmatrix} k\\n \end{pmatrix} \gamma^n(1-\gamma)^{k-n}\ket{k}\bra{k}.
\end{equation}

To evaluate the expected number of detected photons, we expand the POVM elements in the following expression
\begin{align}
    \sum_{n=0}^{\infty}n\hat{\Pi}_n^{\textrm{PNR}} &=\sum_{n=0}^\infty n \sum_{k=n}^\infty \begin{pmatrix}k\\n\end{pmatrix}\gamma^n(1-\gamma)^{k-n}\ket{k}\bra{k}  \\
    &=\sum_{k=0}^{\infty}\ket{k}\bra{k}\sum_{n=1}^{k}k \begin{pmatrix} k-1\\n-1 \end{pmatrix}\gamma^n(1-\gamma)^{k-n}, \nonumber
\end{align}
where the last equality follows from the absorption identity $n\begin{pmatrix}k\\n \end{pmatrix} = k\begin{pmatrix} k-1\\ n-1 \end{pmatrix}$. Substituting $m=n-1$, it follows that
\begin{align}\label{eq:imperfec.}
    \sum_{n=0}^{\infty}n\hat{\Pi}_n^{\textrm{PNR}} &= \sum_{k=0}^{\infty}k\ket{k}\bra{k}\sum_{m=0}^{k-1}  \begin{pmatrix}k-1\\m\end{pmatrix} \gamma^{m+1} (1-\gamma)^{k-1-m} \nonumber \\
    &= \gamma\sum_{k=0}^{\infty} k\ket{k}\bra{k},
\end{align}
where we used the binomial expansion $(x+y)^n=\sum_{k=0}^n \begin{pmatrix}n\\k\end{pmatrix}x^{n-k}y^k$. As a result, $\sum_{n=0}^{\infty}n\hat{\Pi}_n^{\textrm{PNR}}=\gamma \hat{n}$. Therefore,
\begin{equation}
    \sum_{n=0}^{\infty}n\mathrm{Tr}\left[\hat{\rho}_\textrm{B} \hat{D}^\dagger(\beta)\hat{\Pi}_n^{\textrm{PNR}}\hat{D}(\beta)\right]=\gamma\langle\hat{n}\rangle_{\hat{D}(\beta)\hat{\rho}_\textrm{B}\hat{D}^\dagger(\beta)}.
\end{equation}
 
If $\nu\neq 0$, we have
\begin{widetext}
\begin{align}
    \sum_{n=0}^{\infty}n\hat{\Pi}_n^{\textrm{PNR}}&=e^{-\nu}\sum_{k=0}^{\infty}\sum_{l=0}^{\infty}\sum_{n=l}^{k+l}n\frac{\nu^l}{l!} \begin{pmatrix}k\\n-l\end{pmatrix} \gamma^{n-l}(1-\gamma)^{k-(n-l)} \ket{k}\bra{k}\nonumber\\
    &=e^{-\nu} \sum_{k=0}^{\infty} \ket{k}\bra{k} \sum_{l=0}^{\infty}\frac{\nu^l}{l!}\sum_{m=0}^{k}(m+l) \begin{pmatrix}k\\m\end{pmatrix} \gamma^{m}(1-\gamma)^{k-m},
\end{align}
\end{widetext}
where $m=n-l$. Therefore,
\begin{widetext}
\begin{align}\label{eq:dark}
    \nonumber\sum_{n=0}^{\infty}n\hat{\Pi}_n^{\textrm{PNR}}&=e^{-\nu}\sum_{k=0}^{\infty}\ket{k}\bra{k}\sum_{l=0}^{\infty}\frac{\nu^l}{l!}\left[\sum_{m=0}^{k}m \begin{pmatrix}k\\m\end{pmatrix} \gamma^{m}(1-\gamma)^{k-m}+l\sum_{m=0}^{k} \begin{pmatrix}k\\m\end{pmatrix} \gamma^{m}(1-\gamma)^{k-m}\right]\\
    &=\sum_{k=0}^{\infty}\ket{k}\bra{k}\sum_{m=0}^{k}m \begin{pmatrix}k\\m\end{pmatrix} \gamma^{m}(1-\gamma)^{k-m}+ \nu~\mathds{1} ,
\end{align}
\end{widetext}
where in the last line we used the binomial expansion and the identity $\sum_{l=0}^\infty \nu^l/(l-1)!=\nu e^{\nu}$. The first term on the right-hand side of Eq.~\eqref{eq:dark} is associated with the measurement imperfection and does not carry a contribution from dark counts; from Eq.~\eqref{eq:imperfec.} it follows that this term is equal to $\gamma\,\hat{n}$. Therefore, $\sum_{n=0}^{\infty}n\hat{\Pi}_n^{\textrm{PNR}} =\gamma\,\hat{n}+\nu\,\mathds{1}$, and consequently
\begin{equation}
    \sum_{n=0}^{\infty}n\mathrm{Tr}\left[\hat{\rho}_\textrm{B} \hat{D}^\dagger(\beta)\hat{\Pi}_n^{\textrm{PNR}}\hat{D}(\beta)\right]=\gamma\langle\hat{n}\rangle + \nu,
\end{equation}
where the average is taken over the state $\hat{D}(\beta)\hat{\rho}_\textrm{B}\hat{D}^\dagger(\beta)$.

\subsection{Estimation of the transmittance parameter}\label{app:PE2}
 
From Eq.~\eqref{eq:disp_probs_b1}, the ratio between the conditional probabilities is simplified as follows:
\begin{align}
    \frac{p(b_1|a_1)}{p(b_1|a_0)} =\exp\left(\frac{4\gamma \beta\sqrt{\eta}\alpha}{1+\gamma\bar{n}_\textrm{B}}\right),
\end{align}
where the common factors $e^{-\nu}$ and $(1+\gamma\bar{n}_\textrm{B})^{-1}$ cancel in the ratio. Taking $A$ as the logarithm of the above equation, the transmittance estimator follows directly to give Eq.~\eqref{eq:estimator}. From Eq.~\eqref{eq:nestimator}, we find
\begin{align}
(1+\gamma\bar{n}_\textrm{B})A 
&= -\gamma\alpha^2A\eta+\gamma AB,
\end{align}
where $B=(1+\bar{y}-\nu)/\gamma -\beta^2$. Therefore, the estimated value of the parameter $\sqrt{\eta}$ is the solution of the quadratic equation
\begin{align}
    \gamma\alpha^2A\eta + 4\gamma\beta\alpha\sqrt{\eta}-\gamma AB=0.
\end{align}

\bibliography{REFS}
\bibliographystyle{apsrev4-2}

\end{document}